\documentclass[fleqn,usenatbib]{mnras}

\usepackage[T1]{fontenc}
\usepackage{ae,aecompl}

\usepackage{graphicx}   
\usepackage{amsmath}    
\usepackage{amssymb}    

\def\arcmin{\hbox{$^{\prime}$}}

\renewcommand{\vec}[1]{\mbox{\boldmath $#1$}}
\newcommand{\kpc}{{\ \mathrm {kpc}} }
\newcommand{\m}{{\ \mathrm m} }
\newcommand{\s}{{\ \mathrm s} }
\newcommand{\e}{{\ \mathrm e} }
\newcommand{\kg}{{\ \mathrm {kg}} }
\newcommand{\M}{{\ \mathrm {M_\odot}} }
\newcommand{\km}{{\ \mathrm {km}} }
\newcommand{\cm}{{\ \mathrm {cm}} }
\begin{document}

\title[On the radial acceleration of disk galaxies]
{On the radial acceleration of disk galaxies}

\author[Klaus Wilhelm and Bhola N. Dwivedi]{
Klaus Wilhelm,$^{1}$\thanks{E-mail: wilhelm@mps.mpg.de}
Bhola N. Dwivedi $^{2}$\thanks{E-mail: bholadwivedi@gmail.com}
\\
$^{1}$Max-Planck-Institut f\"ur Son\-nen\-sy\-stem\-for\-schung
(MPS), Justus-von-Liebig-Weg 3, 37077 G\"ottingen, Germany\\
$^{2}$Department of Physics, Indian Institute of Technology
(Banaras Hindu University), Varanasi-221005, India}

\date{}

\pubyear{2020}


\label{firstpage}
\pagerange{\pageref{firstpage}--\pageref{lastpage}}
\maketitle

\begin{abstract}
The physical processes defining the dynamics of disk galaxies are still
poorly understood. Hundreds of articles have appeared in the literature over
the last decades without arriving at an understanding within a
consistent gravitational theory. Dark matter (DM) scenarios or a modification
of Newtonian dynamics (MOND) are employed to model the non-Keplerian rotation
curves in most of the studies, but the nature of DM and its interaction with
baryonic matter remains an open question and MOND formulates a mathematical
concept without a physical process. We have continued our attempts to use the
impact theory of gravitation for a description of the peculiar acceleration
and velocity curves and have considered five more galaxies. Using published
data of the galaxies NGC\,3198, NGC\,2403, NGC\,1090, UGC\,3205 and NGC\,1705,
it has been possible to find good fits without DM for the observed disk
velocities and, as example, also for the extraplanar matter of NGC\,3198.
\end{abstract}

\begin{keywords}
galaxies: spiral -- galaxies: kinematics and dynamics -- gravitation --
astroparticle physics
\end{keywords}

\section{Introduction} 
\label{s:introd}

The main topic of this study will be the nearly flat outer
rotation curves (RCs)
of disk galaxies which cannot be understood by the visible baryonic mass
and Keplerian dynamics.

Since \citet{Oor32} and \citet{Zwi33} introduced the concept of
dark matter (DM; \emph{dunkle Materie}) in order to resolve
problems discovered in the velocity distributions of our Galaxy
and in the Coma galaxy cluster, the
DM concept has been applied to explain the flat RCs of
disk galaxies \citep[cf. e.g.][]{Rub83,Rub86}.
Even before, \citet{Bab39} found for the Andromeda Nebula (M 31) a rotation
of the outer portions of the disk with constant
angular velocity indicating that the mass-luminosity coefficient was too
small. He suggested that absorption could be the cause or that
new dynamical considerations might be required.
In the outer parts of the nebula NGC\,3115, \citet{Oor40} also
found that the ratio of mass-density to light-density is
very high.

In a detailed summary of the observations of galaxies and their
interpretations, \citet{Rub00} concludes that galaxy dark halos exist, but,
at the same time, states that details on the amount of DM as well as its
composition and distribution still have to be discovered. Even a modification
of the gravitational potential rather than DM might provide
an explanation for the observations.

\citet{Bur95,TreKoo,McG08,BoBuKa,Weietal,Batetal} and \citet{Sal18}
also conclude that the nature
of DM and its exact role in galaxy formation is still unknown, although it
apparently dominates the dynamics of large-scale cosmic structures.
\citet{Fre08} and \citet{KaReYu} state that the cold dark matter (CDM)
concept is
successful in the Universe, but that there are challenges on galactic scales.
Star formation by bosonic self-interacting dark matter (SIDM)
is considered as possibility \citep{Ebyetal}.
In an investigation of luminous and DM distributions with the help
of extended RCs of 12 disk galaxies \citet{BotPes} conclude that DM halos
have central cores and that the mass ratios between the dark and baryonic
components are between $\approx 9$ to $\approx 5$ for small and large
galaxies, respectively. \citet{Gir00} shows that the DM profiles have a
main component (the ``coupled halo'') and one with a gaslike distribution.
\citet{KrKrGr} find a trend that disks flatten with increased mass, but
conclude that many questions remain open. Many more DM candidates are
mentioned by \citet{Sch17}.

It is thus not surprising
that statements about DM include ``mysterious material'' \citep{BiMaOs},
``DM around galaxies ... as the direct manifestation
of one of its most extraordinary mysteries'' \citep{Donetal,Genetal,KarSal},
mysteries of DM and dark energy \citep{Bek10}. ``Our results may point
to the need for a revision of the current DM paradigm'' \citep{Leletal}.
It is
``... arguably deepest mystery of modern physics'' \citep{Ghaetal};
\citet{deB18} asks whether there is a universal alternative and at the
beginning of their abstract \citet{Rodetal} write:
``Dark matter is currently one of the main mysteries of the
Universe.'' \citet{Sal19} concludes his paper with reference to
the new observational opportunities: ``... if one believes or not
that this will lead to a solution of the old mystery of dark matter.''

A modification of the Newtonian dynamics (MOND) has been proposed
by \citet{Mil83,Mil94,Mil15,Mil16,Mil20}, assuming that for small
accelerations characterized by $a_0 \approx 1.2 \times 10^{-8}\cm\s^{-2}$
the $1/r^2$ dependence in Newton's law of gravity effectively
changes to $1/r$. \citet{GhHaZo} find support that $a_0$ could be a universal
constant. MOND can describe the dynamics
of many galaxies without DM \citep[cf. e.g.][]{KrPaMi,Sam16}.
It has to be noted, though, that it is a mathematical assumption for
which a physical process is still in need \citep[cf. e.g.][]{McG12}.
MOND formulates a fundamental modification of gravity or inertia in
low-acceleration regime -- ``MOND is the key to new gravitational physics''
\citep{SanNoo}.

We have considered the rotation dynamics of
three disk galaxies in a paper entitled
``A physical process of the radial acceleration of disc galaxies''
with a proposal which explains the observations without the need to introduce
any DM \citep{WiDwGal}.
It is our aim to follow up this proposal
by applying the process to five more galaxies with flat or slightly declining
RCs.

\section{Newton's law}
\label{Newton}

\citet{SofRub} hope that the research about galaxies will confirm the
Newtonian gravitational theory or will lead to its successor.
Newton's inverse square law of gravitation is, of course, also a mathematical
abstraction and Newton himself was not convinced that
``action at a distance'' was an appropriate physical process. Since he could
not discover such a process, he formulated his famous ``Hypotheses non
fingo''.
Nicolas Fatio de Duillier influenced by Isaac Newton presented in
1690 his impact theory of gravitation to the Royal Society London.
He had to incorporate a shadow effect to produce an attraction \citep{Fat90};
\citep[cf.][]{Nij01,Bop29,Zeh83}.

Discrepancies between Newton's law and observations, namely, the perihelion
precession of Mercury \citep{Ver59} and the deflection of light twice as
strong as predicted \citep{Dys20,Sol04}, could be resolved in a formal way
by the General Theory of Relativity \citep[GTR,][]{Ein16}.
Later \citet{Lau59}
differentiated between the physical world and its mathematical formulation:
A four-dimensional `world' is only a valuable mathematical trick; deeper
insight, which some people want to see behind it, is not involved.

A physical process based on Nicolas Fatio's idea consistent with GRT\,--\,as
far as the perihelion precession and the light deflection are
concerned\,--\,and in line with strict momentum and energy conservation has
been proposed by \citet{WilWilDwi} and is discussed in a wider context in
\citet{WiDi20}. The solution is based on gravitons
which lose energy and momentum when interacting with baryonic matter.
We have invoked the basic idea of impacting gravitons\,--\,originally called
quadrupoles\,--\,with no mass and a speed of light~$c_0$.
They are absorbed by massive
particles and re-emitted with reduced energy $T_{\rm G}^-$ according to
$T_{\rm G}^- = T_{\rm G}\,(1 - Y)$, where $T_{\rm G}$ is the energy
(very small) of a graviton in the background flux
and $Y$ with $0 < Y \ll1$ is defined as the reduction parameter. The energy
difference $Y\,T_{\rm G}$ leads to a mass increase of the interacting
particle. The corresponding
momentum equation is
$\vec{p}_{\rm G}^- = - \vec{p}_{\rm G}\,(1 - Y)$,
where $|\vec{p}_{\rm G}| = T_{\rm G}/c0$.
This implies that the diminished graviton is
re-emitted in the anti-parallel direction relative to the incoming one.
This geometry had to be assumed in a study defining the mass equivalent
of  the potential energy in a gravitationally bound two-body system.
It is to be noted here that the omni-directional emission,
as originally postulated, led to the conflicts with energy and momentum
conservation \citep{WilDwi15a}.
For small particles and bodies the gravitational interaction obeys the
inverse square law, but in large mass conglomerations
multiple interactions lead to deviations that could explain
the perihelion anomalies \citep{WilDwi}, anomalous Earth flybys
\citep{WilDwi15b} and the non-Keplerian RCs of disk galaxies \citep{WiDwGal}.
Multiple interactions in disk galaxies can only occur within their planes
and are responsible for the flat RCs.
The last topic is also the subject of this paper and will be detailed in
later sections.

The main assumption in this paper is that the flat RCs of
spiral galaxies can be understood without DM. The baryonic mass of galaxies
we denote with $M_{\rm bar}$ and the corresponding velocity and acceleration
quantities with $V_{\rm bar}$ and $g_{\rm bar}$. We follow \citet{KaSaGe}
and others in identifying the ``luminous'' components
of spiral galaxies with the stellar and gaseous disks and a central bulge.
The corresponding velocity, acceleration and mass quantities are
written as $V_{\rm star}$, $V_{\rm gas}$, $V_{\rm bulge}$; $g_{\rm star}$,
$g_{\rm gas}$, $g_{\rm bulge}$; $M_{\rm star}$, $M_{\rm gas}$
and $M_{\rm bulge}$, respectively.
Not all components are present in all spiral galaxies.
The stellar and gaseous parts of the disk are combined to $V_{\rm disk}$,
$g_{\rm disk}$ and $M_{\rm disk}$.
The centripetal acceleration~$g(R)$ exerted by a certain component at
radius~$R$ must cancel the centrifugal force and is thus related
to the rotational velocity~$V$ by
%
\begin{equation}
g(R) = -\frac{V^2}{R}
\label{eqn:centripetal}
\end{equation}
\citep[cf. e.g. Eq. (5)][]{Bab39}.
We use the definitions of \citet{Fre70} and \citet{PeSaSt}:
%
\begin{equation}
\Sigma_{\rm d}(R) = \frac{M_{\rm disk}}{2 \pi R^2_{\rm D}}\e^{-R/R_{\rm D}}~,
\label{eqn:scale_length}
\end{equation}
where $\Sigma_{\rm d}(R)$ is the stellar surface density
and $R_{\rm D}$ the exponential disk scale length.
The radius $R_{\rm gal} = 3.2\,R_{\rm D}$ then encloses 83 \% of the total
light of a galaxy \citep[cf.][]{Sal18}.

Before we present our results, it is
important to attempt a summary of the recent literature on disk galaxies.
It is meant to show the enormous progress that has been achieved in many
aspects, but, at the same time, that fundamental questions are not
resolved\,--\,most of them related to DM.

\section{Spiral galaxies} 
\label{s:spirales}

MOND provides reasonable predictions both for galaxies with low masses and
rotation velocities \citep{MilSan} as well as for massive baryonic systems,
whereas galactic bulges pose a complication \citep{SanNoo}.
Modified gravity thus is an alternative for DM \citep{Bek10,San19}.
\citet{LiTaLi} compare DM models, but cannot prefer one model and exclude
others. Since DM particle have not been found, MOND is the best fit for RCs.
It breaks down, however, for very low
masses of $M_{\rm bar} < 2 \times 10^{36}\kg$
(i.e.~$\approx 10^6\M$)\footnote{Solar mass: $\M$, cf. Table~1 ($^{\rm a}$).}
\citep{Dutetal}.

There is general agreement that a close relationship exists between
baryonic mass and the observed radial acceleration~$g_{\rm obs}$
\citep{McG05a}. It can be described by the
radial acceleration relation (RAR)
%
\begin{equation}
g_{\rm obs} ={\frac{g_{\rm bar}}{1 - \e^{-\sqrt{g_{\rm bar}/g_{\rm t}}}}}
\label{eqn:RAR}
\end{equation}
according to \citet{McLeSc} and \citet{McGetal},
where $g_{\rm t} = 1.2 \times 10^{-10}\m\s^{-2}$.
\citet{Leletal} point out that there is
no room for a substantial variation of $g_{\rm t}$.
Assuming the DM concept, the DM halo contribution is then fully specified by
that of the baryons \citep[see e.g.][]{Tenetal,Ghaetal}.
The authors add that black holes are not important in this context and
\citet{Ricetal} find it remarkable that the tight relationship between DM and
baryonic mass is not affected by the different distributions of the baryons
in galaxies. Many more studies demonstrate that DM halos and stellar disks
are indeed strongly correlated
\citep[e.g.][]{Sanc04,FrSaKa,SalTur,WecTin,Lietala,Lietalb}. This may imply
new physics for DM or a modification of gravity \citep{McG05a}.
Recent observations, related to improved observation
and evaluation techniques \citep[cf. e.g.][]{Behetal}, have revealed very
low halo masses relative to the disk mass adding more constraints
on the galaxy--DM halo connection \citep[cf.][]{Sof16,Posetal}.

The mass discrepancy-acceleration relation (MDAR)
%
\begin{equation}
\frac{V^2_{\rm obs}(R)}{V^2_{\rm bar}(R)} = \mathcal {D}
\label{eqn:MDAR}
\end{equation}
describes a force law in disk galaxies at radius~$R$
\citep{McG14,Leletal,Navetal,Des17}.
\citet{McG08} discusses this relation and MOND and concludes that
the physics behind these empirical relations is unclear.
In the inner regions of dwarf galaxies the baryonic mass distribution cannot
be responsible for the RCs and thus MDAR is not valid there
according to \citet{Sanetal}.

The Tully-Fisher relation (TFR) between luminosity and flat rotation
velocity \citep{TulFis} as well as the corresponding
baryonic Tully-Fisher relation (BTFR) \citep{McG05b}
depend on the fourth power of the flat rotation velocity $V_{\rm f}$
at large $R$:
%
\begin{equation}
M_{\rm bar} = A_{\rm f}\,V_{\rm f}^4,
\label{eqn:BTFR}
\end{equation}
where $A_{\rm f} = 9.94 \times 10^{19}\kg\s^4\m^{-4}$ is the best fit for
the coefficient $A_{\rm f}$ \citep[cf.][]{McGetal}.
It is valid over many decades in mass \citep{McG08} and describes a tight
relation between rotation speed and the luminosity or the baryonic mass
of disk galaxies \citep[cf. e.g.][]{Nav98,Bek10,Leletal},
although its relation to MOND is unclear \citep{McG12}.

Low surface brightness (LSB) galaxies have larger mass discrepancies
between visible and Newtonian dynamical mass than
high surface brightness (HSB) ones \citep{SanNoo,McG14,McGetal}.
We feel that these observations provide important support for our interaction
model as will be discussed in Section~\ref{s:multiple}.

The RC shapes also depend on the surface brightness:
LSB galaxies have slowly rising RCs, whereas in HSB systems the speed
rises sharply and stays flat or even declines
beyond the optical radius \citep{Nav98}.

In the abstract \citet{Nooetal} write:
``At intermediate radii, many RCs decline,
with the asymptotic rotation velocity typically 10 to 20 percent
cent lower than the maximum. The strength of the decline is correlated
with the total luminosity of the galaxies, more luminous galaxies
having on average more strongly declining RCs. At large radii,
however, all declining RCs flatten out, indicating that
substantial amounts of dark matter must be present in these galaxies.''
These findings might be directly relevant for our study,
cf. Sections~\ref{ss.NGC_3198}, \ref{ss.NGC_1090} and \ref{ss.UGC_3205}.

\section{Multiple interactions of gravitons in galactic disks} 
\label{s:multiple}

For isolated masses Newton's law is valid for the impact model, however,
in large mass conglomerations multiple interactions of
the proposed gravitons happen and lead to their multiple energy
decreases. In a spherical configuration, this does not significantly
change the $1/r^2$ dependence of the acceleration field, but in a flat
geometry, such as that of disk galaxies, the multiply affected gravitons
spread out predominantly in the plane of the disk and thus display a
$1/r$ dependence consistent with the flat RCs. It must be noted, however,
that a disk galaxy has a certain thickness, presumably related to its surface
brightness. LSB galaxies will thus, in general, comply with the $1/r$ rule
better than HBS galaxies that are expected to have a $1/r^\beta$ shape with
$\beta > 1$.

We assumed in table~2 of \citet{WiDwGal} that the mean free path length of
gravitons~$\lambda_0$ is equal to $R_{\rm gal}$ and found that an
amplification factor~$F$ of 1.75 to 2 could be achieved by 5 to 6 interactions.
With $\lambda_0 = R_{\rm gal}/4$ the calculation has been extended for this
study and gives a factor of about 5 for 10 iterations.

Details on the disk galaxies NGC\,3198, NGC\,2403,
NGC\,1090, UGC\,3205 and NGC\,1705 will be presented in the following
subsections.

\subsection{NGC\,3198} 
\label{ss.NGC_3198}
\begin{figure}
\includegraphics[width=\columnwidth]{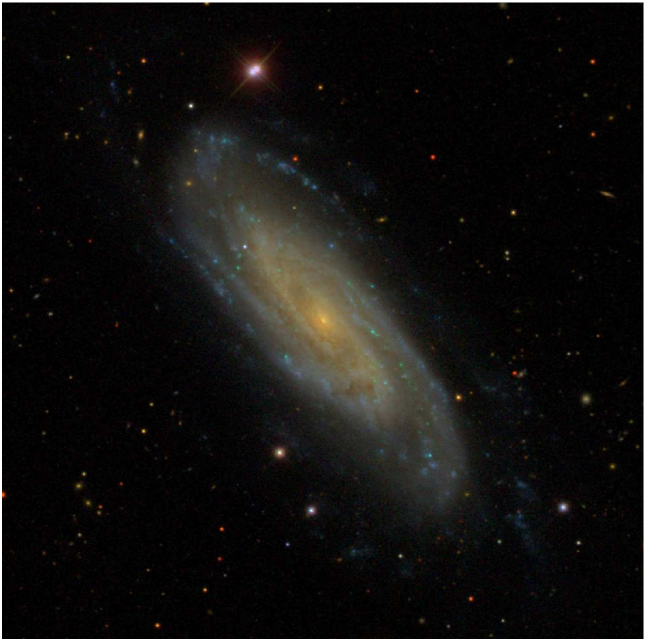}
\caption{Spiral galaxy NGC\,3198.
Type: SB(rs)c. The
apparent size is $8.5\arcmin \times 3.3\arcmin$ \citep{Bra11},
which corresponds to about
$(22.0 \times 8.5)\times 10^{20}\m$ (i.e.~$71.3\kpc \times 27.5\kpc$)
at a distance of 14.4 Mpc \citep{OMea}.
Image:
NGC3198-SDSS-DR14(panorama).jpg.
Credit: Sloan Digital Sky Survey (SDSS), DR14 with SciServer.
}
\label{fig:3198jpg}
\end{figure}

\begin{figure}
\includegraphics[width=\columnwidth]{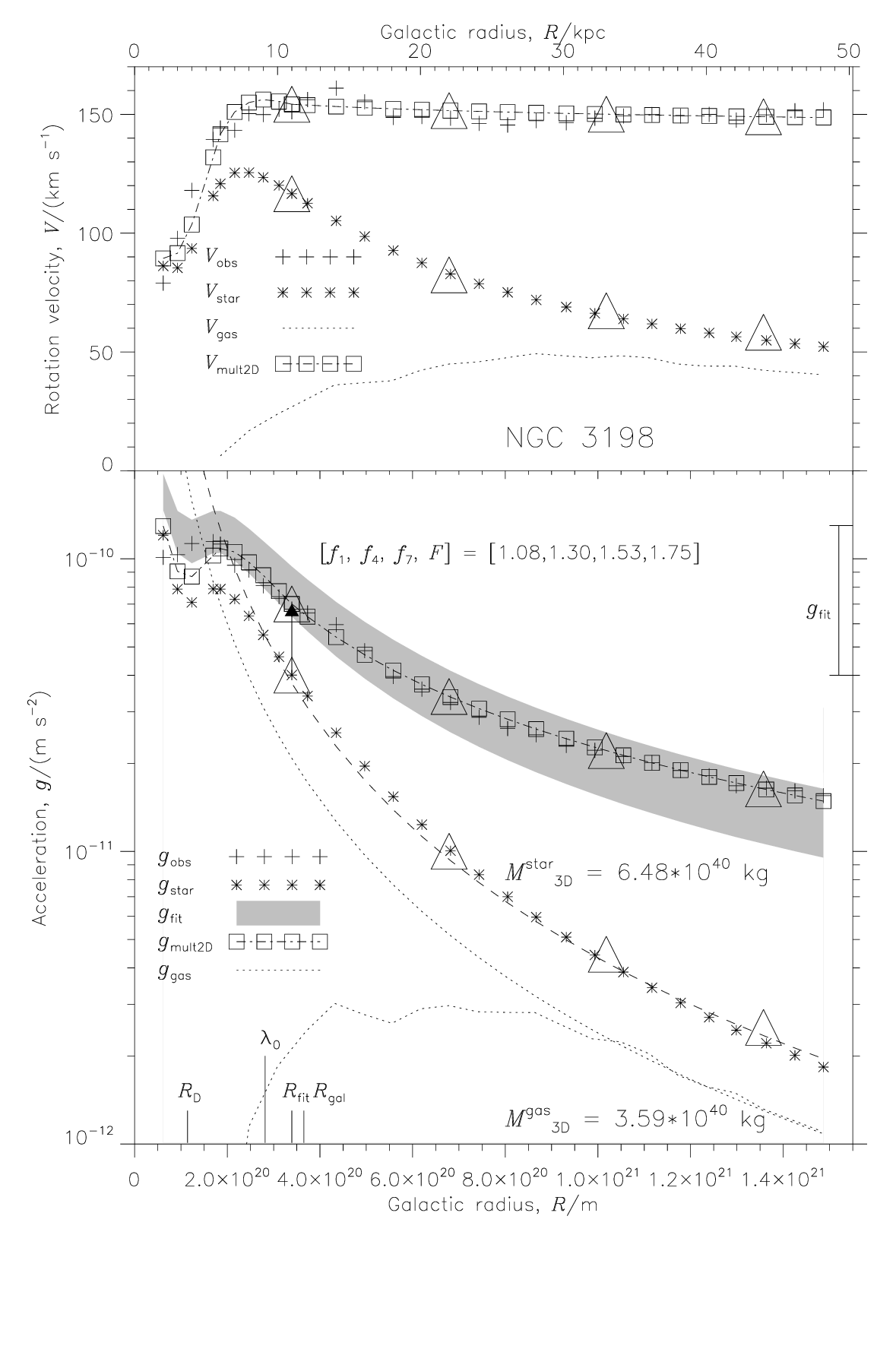}
\vspace{-1.5cm}
\caption{Spiral Galaxy NGC\,3198 (cf. Fig.~\ref{fig:3198jpg}). The $V_{\rm obs}$
and $V_{\rm star}$ data in the upper section are taken from table\,1 of
\citet{KaSaGe} (combined from various references) and $V_{\rm gas}$
from their figure~3. Rotation velocities~$V_{\rm mult2D}$ are calculated
from the acceleration~$g_{\rm mult2D}$ in the lower section
(see Section~\ref{s:multiple}).
The plots~$g_{\rm obs}$, $g_{\rm star}$ and $g_{\rm gas}$ are obtained
with equation~(\ref{eqn:centripetal}) from the corresponding velocities.
The amplification factors~$f_1$ to $F$ and $g_{\rm fit}$
are explained in the text. Three-dimensional fits in the outer reaches of
the galaxy provide estimates on $M_{\rm 3D}^{\rm star}$ and
$M_{\rm 3D}^{\rm gas}$ (cf. Table\,\ref{tab:masses}).
Characteristic scales are:
$R_{\rm D} = 1.14 \times 10^{20}\m$ (i.e.~$3.69\kpc$)
\citep[p. 5,][]{KaSaGe}, $\lambda_0 = 2.82\times~10^{20}\m$
(i.e.~$7.39\kpc$),
$R_{\rm fit} = 3.39 \times 10^{20}\m$ (i.e.~$11.0\kpc$) and
$R_{\rm gal} = 3.65 \times 10^{20}\m$ (i.e.~$11.8\kpc$).
The acceleration~$g_{\rm mult2D}(R)$
is declining with $(R_{\rm fit}/R)^{1.05}$ for $R \ge R_{\rm fit}$.
The superimposed triangles will be explained in Section~\ref{s:discuss}}
\label{fig:3198}
\end{figure}

\begin{figure}
\includegraphics[width=\columnwidth]{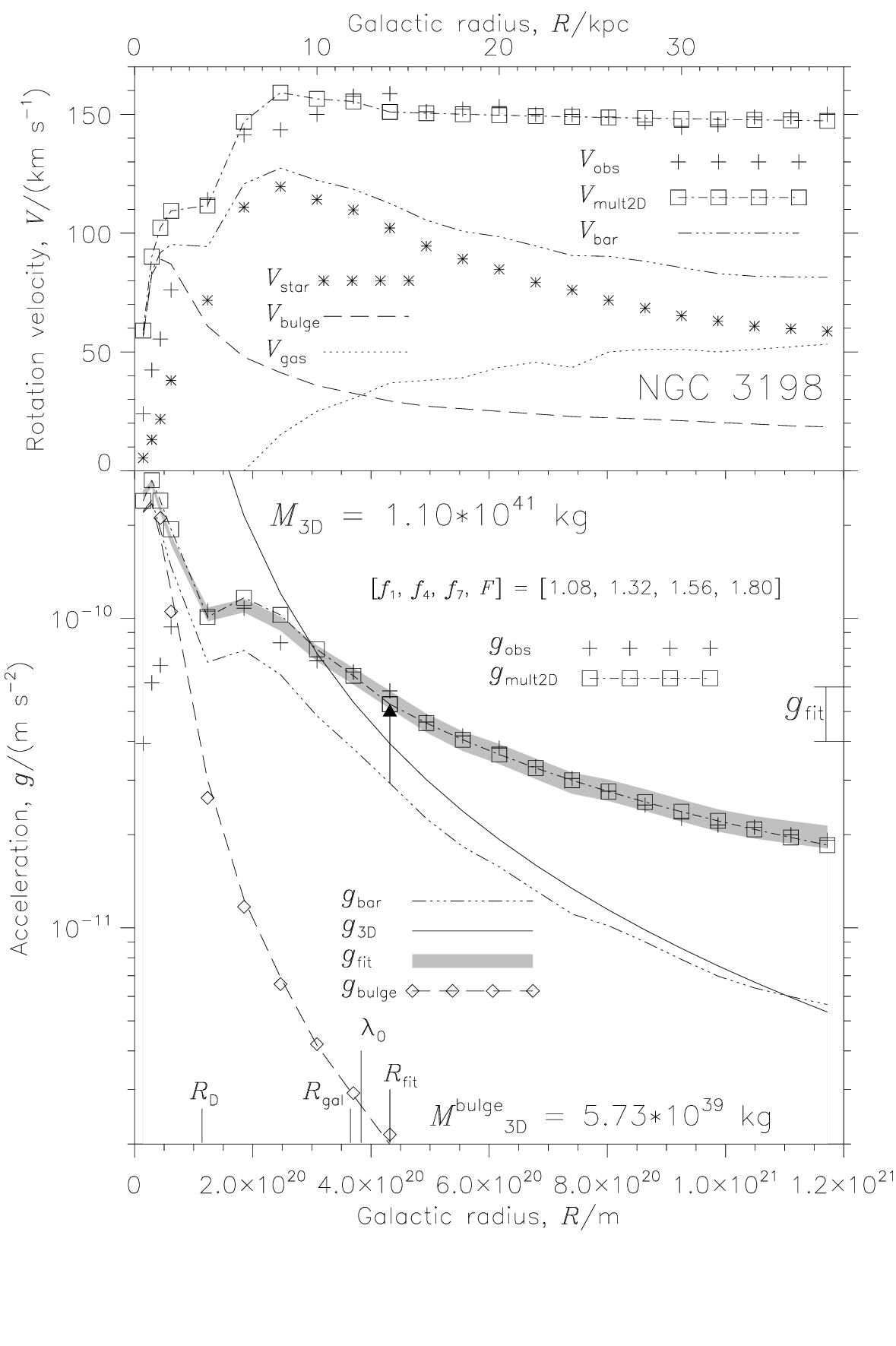}
\vspace{-1.5cm}
\caption{
Spiral Galaxy NGC\,3198 shown with explicit bulge contribution.
The $V_{\rm obs}$, $V_{\rm star}$, $V_{\rm gas}$ and $V_{\rm bulge}$
data in the upper section are taken from figure~37 (ISO, fixed) of
\citet{deB08}. The baryonic rotation velocities in the upper panel
are calculated by
$V_{\rm bar} = (V_{\rm star}^2 + V_{\rm gas}^2 + V_{\rm bulge}^2)^{0.5}$.
$V_{\rm mult2D}$  and the other quantities are obtained
as in Fig.~\ref{fig:3198} and in later diagrams.
The characteristic scales are:
$R_{\rm D}~=~1.14\times~10^{20}\m$ (i.e.~$3.69\kpc$)
\citep[p. 5,][]{KaSaGe},
$\lambda_0~=~3.83\times~10^{20}\m$ (i.e.~$12.4\kpc$),
$R_{\rm fit} = 4.32 \times 10^{20}\m$ (i.e.~$14.0\kpc$) and
$R_{\rm gal}~=~3.65\times~10^{20}\m$ (i.e.~$11.8\kpc$).
The acceleration~$g_{\rm mult2D}(R)$ is
declining with $(R_{\rm fit}/R)^{1.05}$
for $R \ge R_{\rm fit}$.}
\label{fig:Blok}
\end{figure}

The spiral disk galaxy NGC\,3198 is shown in Fig.~\ref{fig:3198jpg}.
\citet{Beg87} observed a large discrepancy between the actual RC and
that predicted from the light distribution in this galaxy.
\citet{KaSaGe} state that it provides spectacular evidence of
a dark force in action as baryons are unable to account for the kinematics.
NGC\,3198 has been studied by many groups
\citep[e.g.][]{Albetal,TaIy,BlAmCa,KadJoWei,Kos06,LovKie,KarSal,AlAmNi,DaoZek}
and we compiled \emph{some} of the results in
Figs.~\ref{fig:3198jpg} to \ref{fig:Blok} as well as in
Table~\ref{tab:masses}.
We did, however, not refer to the many statements
related to DM as we feel that our multiple interaction model can
explain the flat RCs without a DM component.

\begin{figure}
\includegraphics[width=\columnwidth]{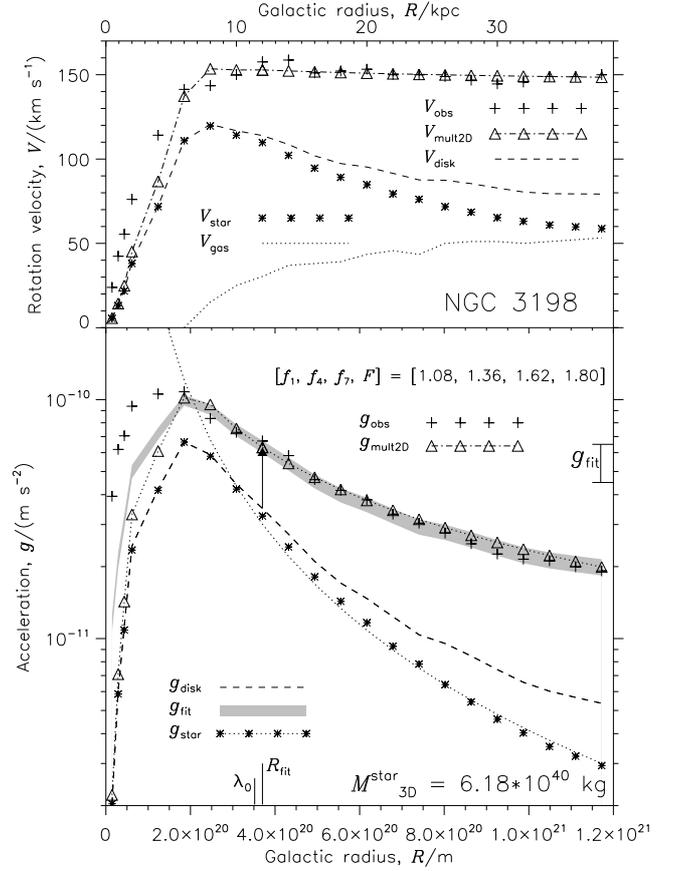}
\vspace{-1.5cm}
\caption{The diagram is the same as in Fig.~\ref{fig:Blok},
but the bulge baryons are not included in
$V_{\rm disk} = (V_{\rm star}^2 + V_{\rm gas}^2)^{0.5}$.
This elimination obviously has a dramatic impact at small $R$,
however, did not influence the RC~$V_{\rm mult2D}$ at large $R$.
The characteristic scales $\lambda_0 = 3.52 \times 10^{20}\m$
(i.e.~$11.4\kpc$) and
$R_{\rm fit} = 3.70 \times 10^{20}\m$ (i.e.~$12.0\kpc$) changed slightly,
but the amplification factor~$F$ and the exponent $\beta$ are not affected.}
\label{fig:Blokstar}
\end{figure}

The RCs published by \citet{KaSaGe} are plotted in the upper panel
of Fig.~\ref{fig:3198} and converted to accelerations in the lower panel
with equation~(\ref{eqn:centripetal}). Using equation~(\ref{eqn:RAR}), i.e.
Eq.~(4) of \citet{McLeSc}, the
curve~$g_{\rm obs}$ was approximated by the shaded area enclosing the
data points. The
required $g_{\rm fit}$ as a modification of $g_{\rm t}$ is given in
Table~\ref{tab:velocities} and the uncertainties are indicated on the
right-hand scale of the diagram. For large radii three-dimensional fits gave
mass estimates of $M_{\rm 3D}^{\rm star}$ and $M_{\rm 3D}^{\rm gas}$.

Under the assumption that most of the baryonic mass is inside
a certain radius $R = \lambda_0$, we can write
%
\begin{equation}
\lambda_0 \approx
\displaystyle{\sqrt{\frac{G_{\rm N}\,M_{\rm bar}}{g_{\rm fit}}}}~,
\label{equ:lambda}
\end{equation}
where $G_{\rm N} = 6.674\,30(15)\times 10^{-11}\m^3\kg^{-1}\s^{-2}$
is the Newtonian constant of gravitation (Source: 2018 CODATA) and
$g_{\rm fit}$ is the acceleration near $\lambda_0$.

On the abscissa of Fig.~\ref{fig:3198}, we have plotted $R_{\rm D}$,
$\lambda_0$ and $R_{\rm gal}$ and notice that at $R_{\rm fit}$ (near
$R_{\rm gal}$) the factor between $g_{\rm star}$ and
$g_{\rm obs}$ is close to 1.75, the value expected, if the mean free
path $\lambda_0$ of the gravitons is comparable to $R_{\rm gal}$.
A near perfect fit to the observations at larger radii is then obtained by
%
\begin{equation}
g_{\rm mult2D}(R) =
g_{\rm mult2D}(R_{\rm fit})\left(\frac{R_{\rm fit}}{R}\right)^\beta
\label{eqn:fit}
\end{equation}
with an exponent~$\beta = 1.05$.
The flat and slightly declining RC in the upper panel,
i.e.~$V_{\rm mult2D}(R)$, then
follows with equation~(\ref{eqn:centripetal}). The physical interpretation
of an exponent~$\beta > 1$ obviously is that the disk of the galaxy has a
certain thickness and the graviton interactions do not exactly occur in
a two-dimensional plane. This conclusion is also supported by the slower
rotation velocity of the extraplanar gas
observed by \citet{Genetal13} as summarized in their conclusions:
``We revealed for the first time in this galaxy the presence of
extraplanar gas over a thickness of a few ($\sim 3$)~kpc. Its amount is
approximately 15 \% of the total mass, and one of its main properties
is that it appears to be rotating more slowly than the gas close
to midplane, with a (rather uncertain) rotation velocity gradient
in the vertical direction (lag) of (7~to~15)$\km\s^{-1}\kpc^{-1}$''.
Other observations of extraplanar gas showing a decrease in velocity above the
galactic plane are mentioned in the next subsection.

In Section~\ref{s:discuss} we will discuss this configuration in more detail
taking into account energy and angular momentum conservation.

The acceleration~$g_{\rm mult2D}(R_{\rm fit})$ also provides a starting
point for approximations at smaller radii. The multiple interaction scenario
implies that inside the disk gravitons travel back and forth after multiple
energy reductions. The amplification factor should, therefore, be close to
zero at $R = 0$. A linear interpolation between $R_{\rm fit}$ and $R = 0$
gives a very good fit. The values of four of the factors are plotted in the
diagram.

The irregular behaviour of $g_{\rm obs}$ near the centre of NGC\,3198
indicates the presence of a bulge. In Fig.~\ref{fig:Blok} this component
is explicitly included with data of \citet{deB08}. The treatment of the
data is very similar to that of Fig.~\ref{fig:3198}. As a result only
$\lambda_0$ and $R_{\rm fit}$ are shifted to somewhat larger values and
$g_{\rm fit}$ decreased, but still is close to the acceleration, where
the amplification factor~$F$ is 1.75 at $R_{\rm fit}$.

Since the increased accelerations appear to be directly related to the
disk-like distribution of the baryonic matter, the more spheroidal bulge
component should not significantly contribute to the effect. In order to test
this expectation, we eliminated in Fig.~\ref{fig:Blokstar} the bulge
before calculating $V_{\rm disk}(R)$. This obviously had a major effect
on the rotation speeds and accelerations at small radii, however, had nearly
no influence on $V_{\rm mult2D}(R)$ for $R \ge R_{\rm fit}$.

\subsection{NGC\,2403} 
\label{ss.NGC_2403}

\begin{figure}
\begin{center}
\includegraphics[width=0.9\columnwidth]{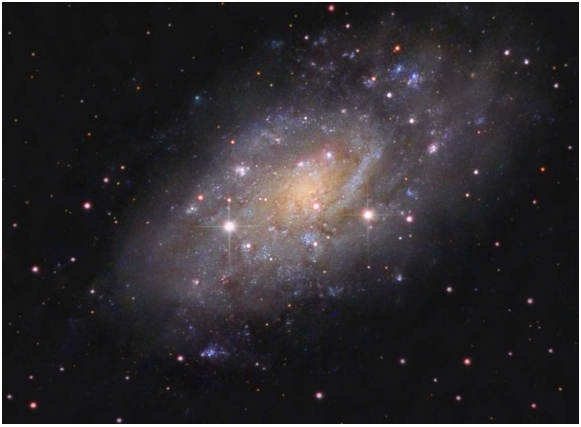}
\end{center}
\caption{
Spiral galaxy NGC\,2403.
Type: SAB(s)cd,
apparent size $21.9\arcmin \times 12.3\arcmin$,
which corresponds to about
$(10.6\times~5.9)\times~10^{20}\m$ (i.e.~$34.3\kpc \times 19.1\kpc$)
at a distance of 2.69~Mpc
(NASA/IPAC Extragalactic Database).
Credit and Copyright Adam Block/Mount Lemmon SkyCenter/University of
Arizona - www.caelumobservatory.com/gallery/n2403.shtml.}
\label{fig:2403jpg}
\end{figure}

The spiral galaxy NGC\,2403 is shown in Fig.~\ref{fig:2403jpg}.
This LSB galaxy is also the subject of many publications
\citep[e.g.][]{Jaretal,LovKie,McG14}.

%
%
\begin{figure}
\includegraphics[width=\columnwidth]{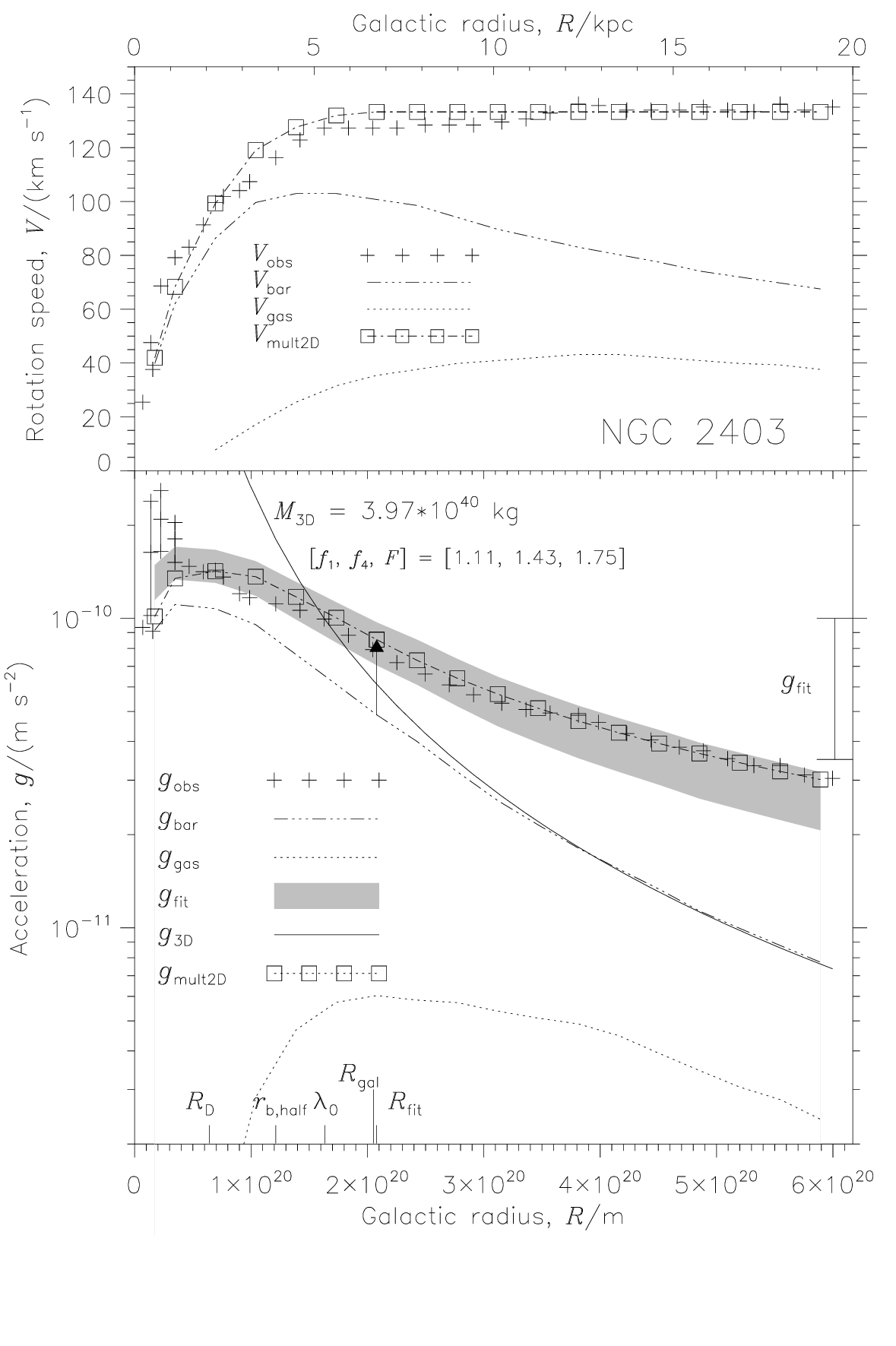}
\vspace{-1.5cm}
\caption{
Spiral Galaxy NGC\,2403 (cf. Fig.~\ref{fig:2403jpg}).
The $V_{\rm obs}$, $V_{\rm bar}$ and $V_{\rm gas}$ velocity data are taken
from figure~1 of \citet{FriMSal}.
Characteristic scales are:
$R_{\rm D}~=~6.42\times~10^{19}\m$ (i.e.~$2.08\kpc$)
\citep{McG05b},
$r_{\rm b,half} = 1.21 \times 10^{20}\m$ (i.e.~$3.92\kpc$) \citep{Sanetal},
$\lambda_0 = 1.64 \times 10^{20}\m$ (i.e.~$5.31\kpc$),
$R_{\rm gal} = 2.05 \times 10^{20}\m$ (i.e.~$6.64\kpc$) and
$R_{\rm fit}~=~2.08\times~10^{20}\m$ (i.e.~$6.74\kpc$).
The flat RC is not declining, i.e.~$\beta \approx 1$.}
\label{fig:2403}
\end{figure}
%
\begin{figure}
\begin{center}
\includegraphics[width=\columnwidth]{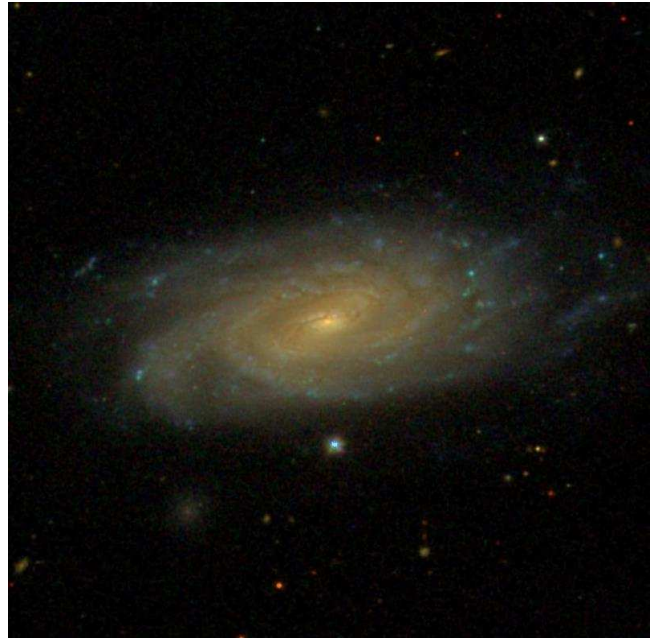}
\end{center}
\caption{
Spiral galaxy NGC\,1090.
Type: SB(rs)bc,
apparent size $3.9\arcmin \times 1.8\arcmin$ ,
which corresponds to about
$(2.6 \times 1.2) \times 10^{21}\m$  (i.e.~$84.3\kpc \times 38.9\kpc$)
at a distance of 37.5 Mpc \citep{Bra11}.
Image:
NGC1090-SDSS-DR14.jpg.
Credit: Sloan Digital Sky Survey (SDSS), DR14.}
\label{fig:1090jpg}
\end{figure}
%
\begin{figure}
\includegraphics[width=\columnwidth]{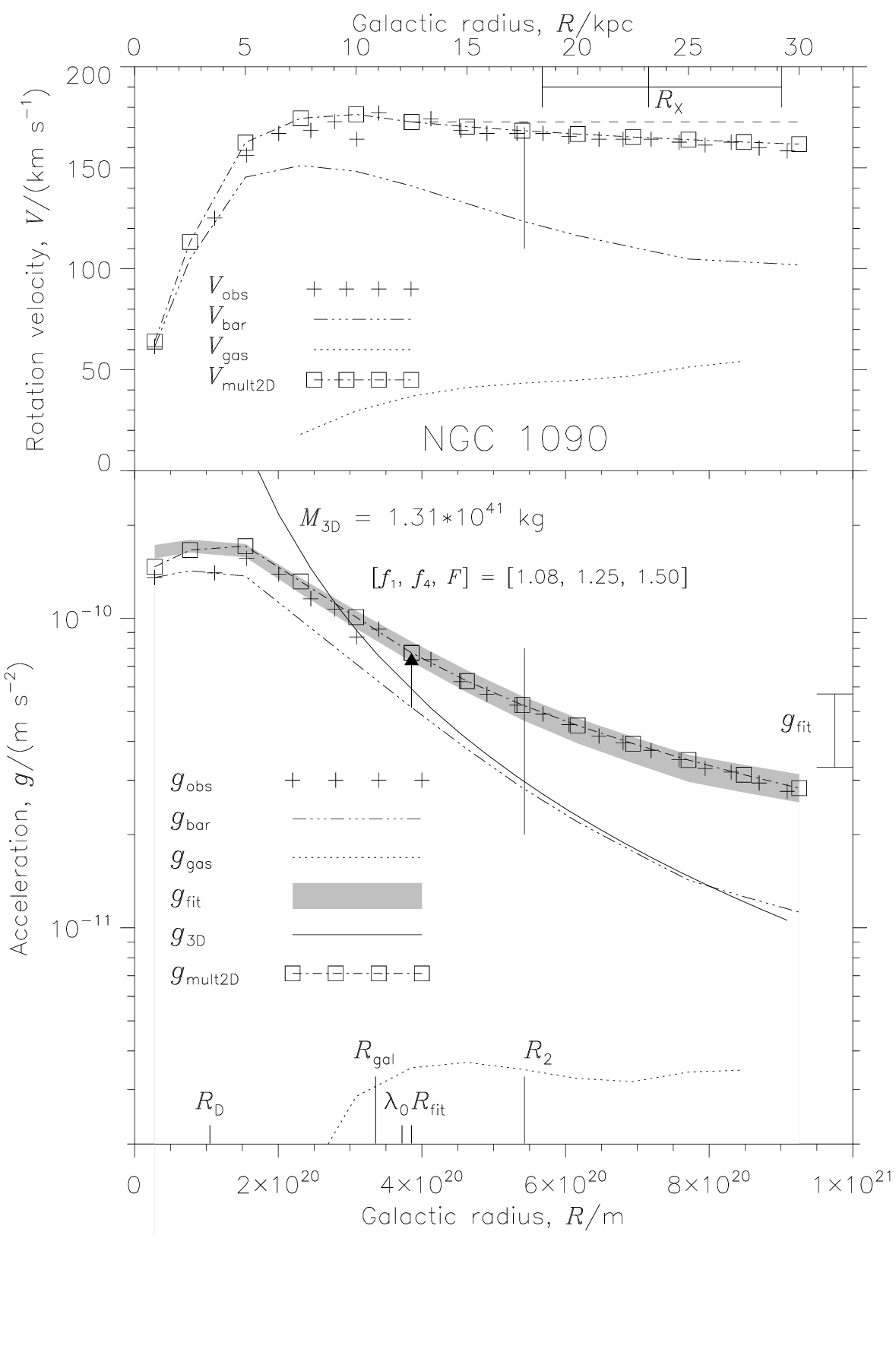}
\vspace{-1.5cm}
\caption{
Spiral Galaxy NGC\,1090 (cf. Fig.~\ref{fig:1090jpg}).
The $V_{\rm obs}$, $V_{\rm bar}$ and $V_{\rm gas}$ data are taken from
figure~1 of \citet{FriMSal}. The characteristic scales are:
$R_{\rm D}~=~1.05\times~10^{20}\m$ (i.e.~$3.40\kpc$) \citep{FriMSal},
$R_{\rm gal}~=~3.36\times~10^{20}\m$ (i.e.~$10.9\kpc$),
$\lambda_0 = 3.72 \times 10^{20}\m$ (i.e.~$12.1\kpc$),
$R_{\rm fit}~=~3.86\times~10^{20}\m$ (i.e.~$12.5\kpc$) and
$R_{\rm X}~=~23.3\kpc$) with 0.1 dex uncertainties \citep[see table 2 in
arXiv:astro-ph/0602027v1,][]{KadJoWei}.
The amplification factor at $R_{\rm fit}$ is $F = 1.50$, whereas
a value of 1.75 was found for NGC~3198 and
NGC~2403. The
acceleration~$g_{\rm mult2D}(R)$ is declining with $(R_{\rm fit}/R)^{1.15}$
for $R \ge R_{\rm fit}$. The dashed line in the upper panel indicates a
constant flat RC. $R_2$~will be discussed in the text.}
\label{fig:1090}
\end{figure}
%
\begin{figure}
\includegraphics[width=\columnwidth]{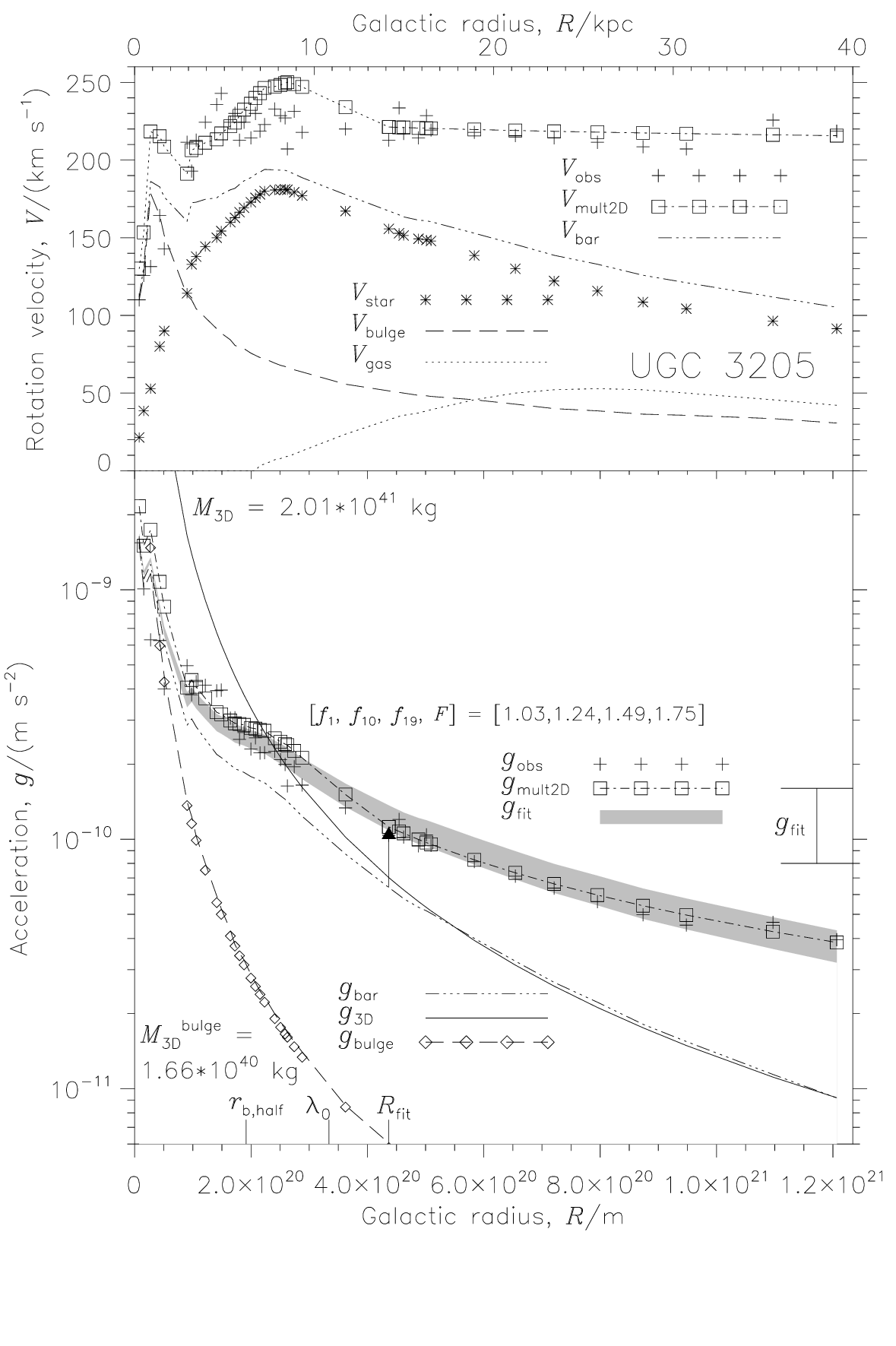}
\vspace{-1.5cm}
\caption{Spiral Galaxy UGC~3205 of Type Sab.
The velocities $V_{\rm obs}$, $V_{\rm star}$, $V_{\rm gas}$ and $V_{\rm bulge}$
data in the upper section are taken from figure~1 of
\citet{SanNoo} as well as ``the Newtonian sum''
$V_{\rm bar} = (V_{\rm star}^2 + V_{\rm gas}^2 + V_{\rm bulge}^2)^{0.5}$.
Characteristic scales are:
$r_{\rm b,half} = 1.92 \times 10^{20}\m$ (i.e.~$6.22\kpc$) \citep{Sanetal},
$\lambda_0 = 3.34 \times 10^{20}\m$ (i.e.~$10.8\kpc$) and
$R_{\rm fit}~=~4.37\times~10^{20}\m$ (i.e.~$14.2\kpc$).
The acceleration~$g_{\rm mult2D}(R)$ is declining with $(R_{\rm fit}/R)^{1.05}$
for $R \ge R_{\rm fit}$, resulting in $\Delta V_{\rm mult2D} \approx 6$~km/s
at $R$ = 40 kpc relative to a flat~RC.}
\label{fig:3205bar}
\end{figure}
%
\begin{figure}
\includegraphics[width=\columnwidth]{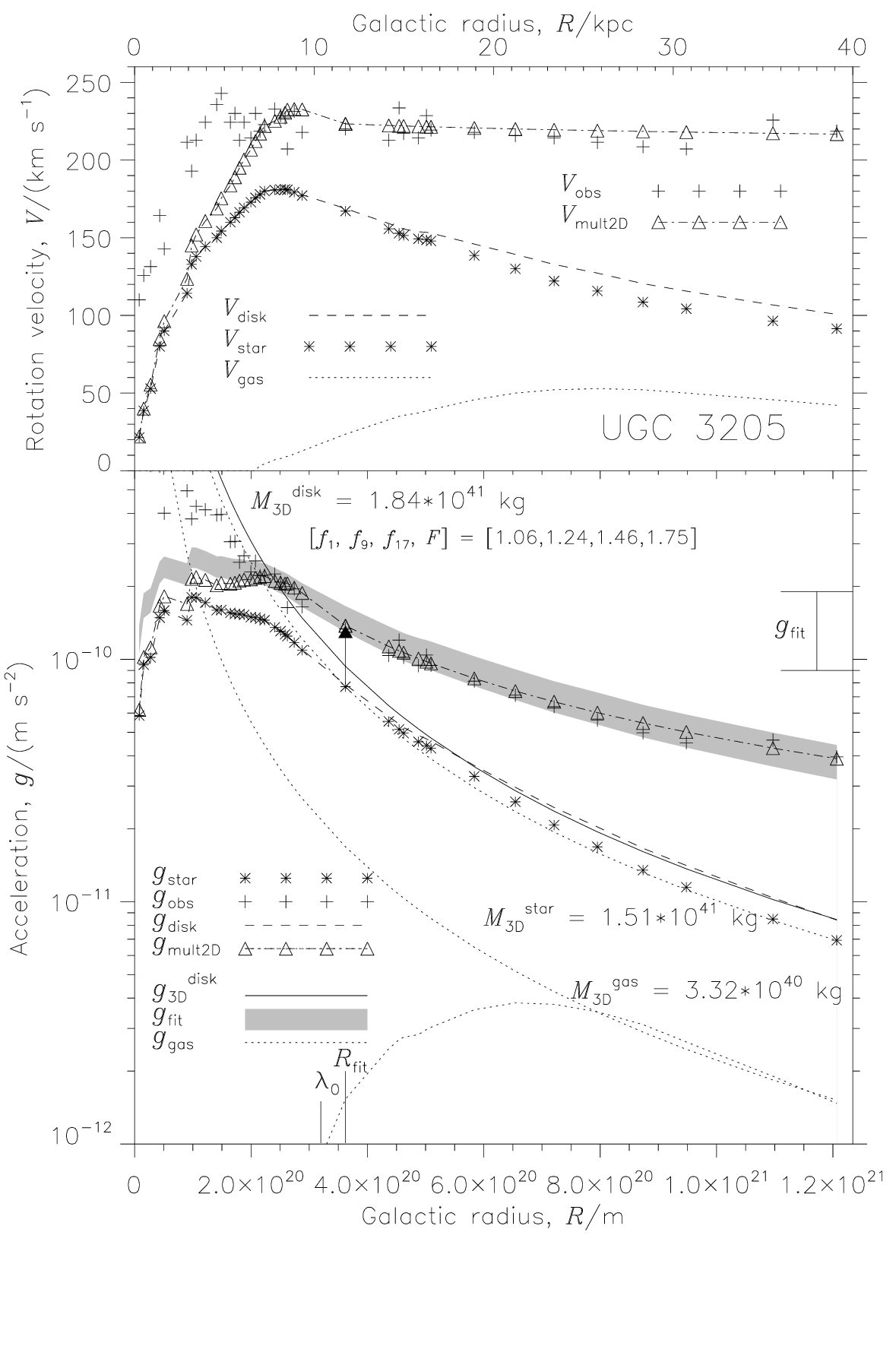}
\vspace{-1.5cm}
\caption{These diagrams are based on the same data as in
Fig.~\ref{fig:3205bar} \citep[][figure~1]{SanNoo},
but the bulge component is not included. An important result of the
$V_{\rm mult2D}$ and $g_{\rm mult2D}$ calculations is that the elimination of
the bulge baryons does not effect the RC at large radii.
The characteristic scales
$\lambda_0 = 3.20 \times 10^{20}\m$ (i.e.~$10.4\kpc$) and
$R_{\rm fit} = 3.63 \times 10^{20}\m$ (i.e.~$11.8\kpc$)
changed only very little,
but the amplification factor~$F$ and $\beta$ remain the same.}
\label{fig:3205disk}
\end{figure}
%
\begin{figure}
\includegraphics[width=\columnwidth]{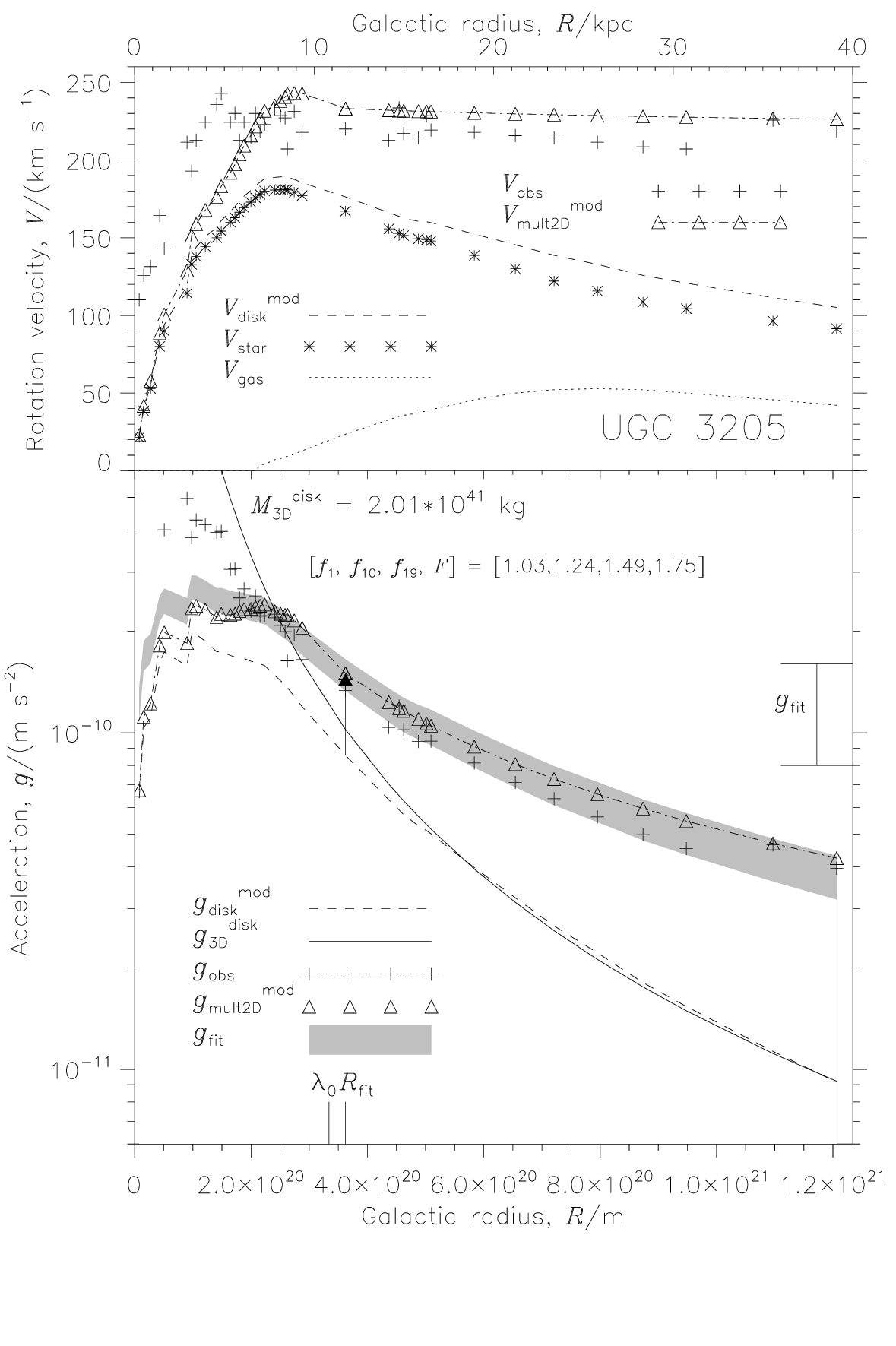}
\vspace{-1.5cm}
\caption{
These diagrams are again based on the same data as in
Fig.~\ref{fig:3205bar} taken
from figure~1 of \citet{SanNoo},
but the bulge component is now distributed over the disk by assuming
a modified
$g_{\rm disk}^{\rm mod} =
g_{\rm disk} \times M_{\rm disk}^{\rm mod}/M_{\rm disk}$
(cf. Table\,\ref{tab:masses}).
This has a major impact on the extended RC with
$V_{\rm mult2D}^{\rm mod}$ higher than $V_{\rm mult2D}$
by more than $10~\km\s^{-1}$.
The characteristic scales are
$\lambda_0 = 3.34 \times 10^{20}\m$  (i.e.~$10.8\kpc$) and
$R_{\rm fit} = 3.62 \times 10^{20}\m$ (i.e.~$11.7\kpc$).}
\label{fig:3205DB}
\end{figure}
%
\begin{figure}
\begin{center}
\includegraphics[width=0.6\columnwidth]{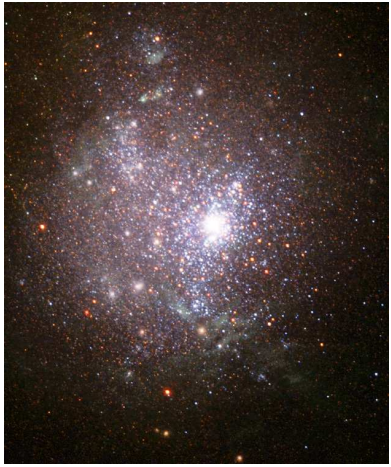}
\end{center}
\caption{Irregular dwarf galaxy NGC\,1705, Type: SAO - pec.
Apparent size: $1.9\arcmin \times 1.4\arcmin$ (Wikipedia),
which corresponds to about
$(1.8 \times 1.3)\times10^{20}\m$ (i.e.~$5.83\kpc \times 4.21\kpc$)
at a distance of $(5.1 \pm 0.6)$~Mpc
\citep{OHaetal}.
Credit:
NASA, ESA and the Hubble Heritage Team (STScI/AURA),
Acknowledgment: M. Tosi (INAF, Osservatorio Astronomico di Bologna)}
\label{fig:1705jpg}
\end{figure}
%
\begin{figure}
\includegraphics[width=\columnwidth]{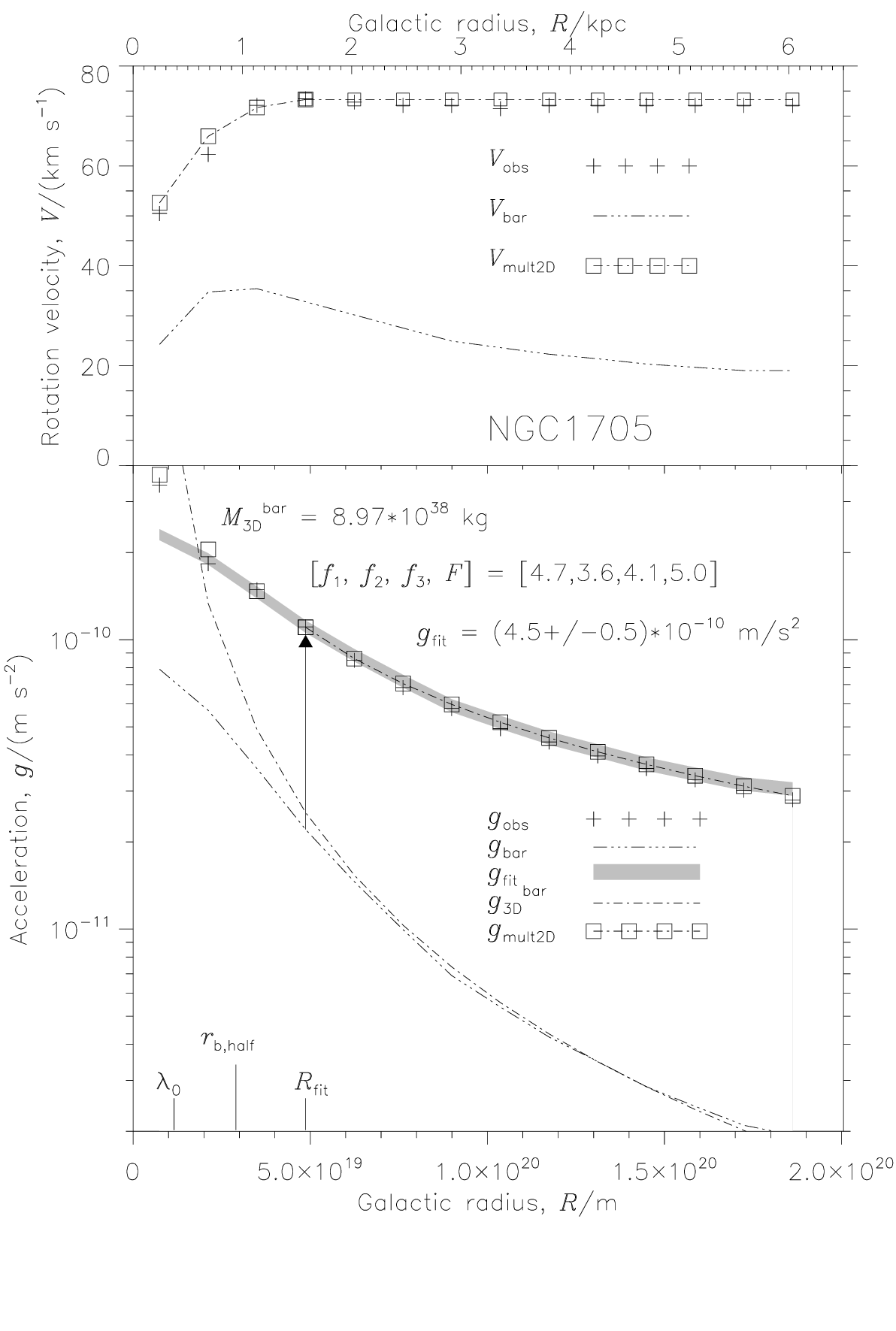}
\vspace{-1.5cm}
\caption{
Spiral Galaxy NGC\,1705
(cf. Fig.~\ref{fig:1705jpg}). The $V_{\rm obs}$
and $V_{\rm bar}$ data in the upper section are taken from figure~5 of
\citet{Sanetal}.
Characteristic scales are:
$\lambda_0 = 1.15 \times 10^{19}\m$ (i.e.~$0.373\kpc$),
$r_{\rm b,half} = 2.90 \times 10^{19}\m$ (i.e.~$0.940\kpc$)
and $R_{\rm fit} = 4.87 \times 10^{19}\m$ (i.e.~$1.58\kpc$). Note the large
amplification factor~$F = 5$ and $\beta \approx 1$.}
\label{fig:1705}
\end{figure}

It is discussed by \citet{Jaetal11} in the context of ``a global thin-disc model
of spiral galaxies'' as a system with vertical gradients of the azimuthal
velocity. Based on data from \citet{Fraetal02}, they find
a near linear decrease at heights
above the plane between (0.6 and 3.0) kpc
of $(- 10 \pm 4)\km\s^{-1}\kpc^{-1}$. This decrease is consistent with
other observations of peculiar kinematics
\citep{SchSS,Fraetal,FraOosSan,FraBin08,Fra10},
but \citet{FraBin06} point out
that their model does not predict the observed inflow.

In Fig.~\ref{fig:2403} the RCs taken from figure~1 of \citet{FriMSal}
display a perfectly flat $V_{\rm obs}$ that could best be fitted with
$\beta \approx 1$. For a thin, LSB galaxy such a result could be expected as
most of the graviton interactions have to occur in the plane of the disk.
Outside this plane, the extraplanar gas will experience normal
gravitational attraction by $M_{\rm bar}$ with a Keplerian rotation
velocity, i.e. slower than the $V_{\rm disk}$.

\subsection{NGC\,1090} 
\label{ss.NGC_1090}

In Fig.~\ref{fig:1090} we have modelled the observed RC of spiral galaxy
NGC\,1090 shown in Fig.~\ref{fig:1090jpg} quite successfully as
$V_{\rm mult2D}(R)$ under the assumption of an amplification factor~$F = 1.5$
and an exponent~$\beta = 1.15$. In this case, the high exponent might be
caused by the small amplification factor~$F$ and a relatively
great $\lambda_0$ indicating fewer multiple interactions of
gravitons.

\citet{LovKie} write that there is a clear need for
CDM for explaining the RCs of NGC\,1090
(and other galaxies), however, its distribution is unclear.
\citet{KadJoWei} define a radius~$R_{\rm X}$, where the relative contribution
of DM to the observed RC $V_{\rm obs}$ equals that of the baryonic
mass~$M_{\rm bar}$. For  NGC\,1090 they find $R_{\rm X} = 23.2\kpc$ and
emphasize the findings of e.g. \citet{PerSal} and \citet{PeSaSt} that for
LSB galaxies, such as NGC\,1090 as extreme case, cf. \citep{Geetal04}
and also http://www.kopernik.org/images/archive/n1090.htm,
the RCs deviate from the baryonic contribution at smaller radii
than for HSB ones. The radius~$R_2$,
where the value of $g_{\rm mult2D}$ is twice that of $g_{\rm bar}$,
can be compared to $R_{\rm X}$ indicating that the multiple interaction
process can explain the apparent mass discrepancy.

\subsection{UGC\,3205} 
\label{ss.UGC_3205}

In Figs.~\ref{fig:3205bar} to \ref{fig:3205DB}, we plot RCs of the
early-type (HSB) disk galaxy UGC\,3205 with data taken from \citet{SanNoo},
which are slightly modified in the next two figures.
In Fig.~\ref{fig:3205disk} the bulge component is not included and
in Fig.~\ref{fig:3205DB} the bulge component is distributed over the disk.

The observed RC is declining
at large radii. In this context, it may be of interest that
many galaxies with declining RCs are found at $z > 1$, interpreted as
possibly indicating smaller DM contributions \citep[e.g.][]{Genetal17}.
\citet{Dreetal}, however, published a counterexample.
\citet{SanNoo} present a fit of UGC\,3205 with MOND and
find deviations from the observed velocities below 10~kpc,
tentatively attributed to the weak bar of this galaxy, but
quite good agreement at large radii.

We apply our multiple interaction model in Fig.~\ref{fig:3205bar} and
obtain a good fit even at small radii. This can be seen, in particular,
in the acceleration diagram in the lower panel.
The similarity between the distribution of the bulge component of this
galaxy with that of NGC\,3198 in Fig.~\ref{fig:Blok} motivated us to
apply the same
procedure of eliminating the bulge contribution in Fig.~\ref{fig:3205disk}.
The result again is that the presumably spheroidal bulge contribution
does not affect the RC at large radii.
The claim of \citet{McG14} that the BTFR
is insensitive to the distribution of the baryonic mass in a galaxy, has
been tested in Fig.~\ref{fig:3205DB}, where the bulge mass has been
distributed over the disk. This has a major impact on the extended RC.
The relative increase of the resulting
$V^{\rm mod}_{\rm mult2D}$ is greater than 5 \% and thus would give
a fractional gain in baryonic mass of more than 20 \% in
equation~({\ref{eqn:BTFR}}).
\citet{Sanetal} also found that the RCs are affected by the distribution
of the baryonic mass.

\subsection{NGC\,1705} 
\label{ss.NGC_1705}

The four galaxies discussed in the preceding sections are all characterized
by regular disk shapes. The irregular dwarf galaxy NGC\,1705 we want to
include in our sample in order to test the applicability of our procedure
on a broader scale. This compact galaxy is much smaller and contains in its
nucleus a luminous super star cluster with a mass of
$\approx 2~\times~10^{35}\kg$ (i.e.~$10^5\M$)
and a significant cold dust component
\citep{Tosetal,Annetal,OHaetal}.

The RCs in Fig.~\ref{fig:1705} demonstrate that there is a very large
discrepancy between $V_{\rm bar}$ and $V_{\rm obs}$. A conclusion
of \citet{Sanetal} on the shapes of dwarf galaxy RCs is\,--\,stated
explicitly for NGC\,1705\,--\,that they are difficult to reproduce with DM
scenarios, if the central baryon component is small.

If we use the same procedure for this galaxy as for the other ones, we
find a very large $g_{\rm fit}$ and a small $\lambda_0$ relative
to~2~$r_{\rm b,half}$. As mentioned in Sect.~\ref{s:multiple} this configuration
allows many multiple interactions and produces an amplification factor of 5
for 10 iterations. This factor gives a good fit for
$R > R_{\rm fit} \approx 2~r_{\rm b,half}$. Inside $R_{\rm fit}$ the
amplification factor has to remain high down to $\lambda_0$. The data
do, however, not allow any conclusion for $R < \lambda_0$.

\section{summary}
\label{s:summary}

%
\begin{table*}
\centering
\caption{Baryonic mass values of the five disk galaxies from the literature
(without uncertainty margins) and from fits in the figures.}
\begin{tabular}{|l|l|c|c|c|c|c|c|c|} \hline
&Mass($^{\rm a}$)& $M_{\rm bulge}$/ & $M_{\rm star}$/ & $M_{\rm gas}$/ &
$M_{\rm disk}$/ & $M_{\rm bar}$/ & $M_{\rm 3D}$/ & \\
Galaxies & Figures & $10^{40}$\,kg & $10^{40}$\,kg & $10^{40}$\,kg &
$10^{40}$\,kg & $10^{40}$\,kg & $10^{40}$\,kg & References\\
\hline
NGC\,3198 &&& 6.76 & 0.994 &&&& \citet{Beg89}\\
&&&&& 12.9 &&&\citet{Kos06}\\
&&& 2.62 &3.88 &&&&\citet{StMcSw}\\
&Fig.~\ref{fig:3198}&&(6.48)($^{\rm b}$)&(3.59) &8.75& &(10.1)&\citet{KaSaGe}\\
& Fig.~\ref{fig:Blok} & 0.573 & 5.60 &&&  &(11.0) & \citet{deB08}, figure~37\\
& Fig.~\ref{fig:Blokstar}&\{ - \}($^{\rm c}$) &(6.18)&&&&&
\\
\hline
NGC\,2403 & Fig.~\ref{fig:2403}&
&& 0.887 & 2.41 & [3.29]($^{\rm d}$) & (3.97)&\citet{FriMSal} \\
&&& 2.19 & 0.934 & [3.12] & [3.12] && \citet{McG05b} \\
&&&&& & 1.84&& \citet{Sanetal} \\
&& 2.41 & 0.887 &0.801 &&& 3.97 & \citet{deBetal} \\
&&  & &1.04&&&  & \citet{Fraetal02} \\
&&&0.810 &1.17 &&&&\citet{StMcSw}\\
\hline
NGC\,1090 & Fig.~\ref{fig:1090} &
&& [1.68]($^{\rm e}$) & 9.34 & [11.3] & (13.1) & \citet{FriMSal}  \\
&&&&  1.69 &&&& \citet{Geetal04}  \\
&&&&&&10.7&& \citet{KadJoWei}  \\
\hline
UGC\,3205&&&&&& 13.9 && \citet{Sanetal} \\
& Fig.~\ref{fig:3205bar} &
(1.66) &&&  && (20.1) & \\
& Fig.~\ref{fig:3205disk} &
\{ - \}($^{\rm c}$)& (15.1) & (3.32)&(18.4) &  &  & \\
& Fig.~\ref{fig:3205DB} &
\{ - \}($^{\rm f}$) &&& [20.1] &&  & \\
\hline
NGC\,1705 &&&$\ge 0.0557$ &$\approx 0.0239$ &&&&\citet{Annetal}  \\
& Fig.~\ref{fig:1705} &
&&&& 0.0897 & (0.0897) &\citet{Sanetal}  \\
\hline
\end{tabular} \\
($^{\rm a}$): Mass of Sun $(1.98847 \pm 0.00007) \times 10^{30}\kg = 1\M$
(International Astronomical Union);\\
($^{\rm b}$): Results of three-dimensional fits in figures are given
in parentheses;
($^{\rm c}$): $M_{\rm bulge}$ neglected;\\
($^{\rm d}$): Values in brackets with modifications explained in text;
($^{\rm e}$): 18~\% of $M_{\rm bar}$ \citep{FriMSal};\\
($^{\rm f}$): $M_{\rm bulge}$ distributed over the disk.

\label{tab:masses}
\end{table*}
%
\begin{table*}
\centering
\caption{Observed rotation velocities of the five disk galaxies
together with derived mass and acceleration values.}
\begin{tabular}{|l|l|c|c|c|c|c|c|c|c|} \hline
Galaxy &  &
$V^{\rm max}_{\rm obs}$/ & $V^{\rm max}_{\rm f}$/ &
$A_{\rm fit}$/~~~~~($^{\rm a})$
& [$M_{\rm fit},M_{\rm f}$]/&
$V^{\rm min}_{\rm obs}$/ & $g^{\rm min}_{\rm mult2D}$/
& $g_{\rm fit}$/ &$\beta$($^{\rm b}$)\\
&Figures & $\km\s^{-1}$ & $\km\s^{-1}$ &  $\kg\s^4\m^{-4}$ & $10^{40}\kg$
&$\km\s^{-1}$
&$\m\s^{-2}$&$\m\s^{-2}$ &  \\
\hline
NGC\,3198 & Fig.~\ref{fig:3198}&
161.0 & 154.2 & $1.76\times10^{20}$ & $[9.98,5.63]$ & 148.7
& $1.49\times10^{-11}$ & $8.50\times10^{-11}$ & 1.05\\
&Fig.~\ref{fig:Blok}&
158.7 & 151.0 & $3.00\times10^{20}$ & $[15.6,5.17]$ & 147.3 &
$1.85\times10^{-11}$ & $5.00\times10^{-11}$& 1.05\\
& Fig.~\ref{fig:Blokstar}&
158.7 & 152.8 & $2.72\times10^{20}$ & $[14.9,5.42]$ & 148.5 &
$1.88\times10^{-11}$ & $5.50\times10^{-11}$& 1.05 \\
\hline
NGC\,2403&Fig.~\ref{fig:2403}&
136.2 & 133.3 & $2.50\times10^{20}$ & $[7.88,3.13]$ & 133.3 & $3.01\times10^{-11}$
& $6.00\times10^{-11}$& $\approx 1.00$\\
NGC\,1090&Fig.~\ref{fig:1090}&
177.2 & 172.7 & $3.33\times10^{20}$ & $[29.6,8.84]$ & 161.7 & $2.83\times10^{-11}$
& $4.50\times10^{-11}$& 1.15\\
\hline
UGC\,3205& Fig.~\ref{fig:3205bar}&
242.9 & 221.2 & $1.25\times10^{20}$ &$[29.9,23.8]$ & 215.7 & $3.86\times10^{-11}$
& $1.20\times10^{-10}$& 1.05 \\
&Fig.~\ref{fig:3205disk}&
242.9 & 223.3 & $1.25\times10^{20}$ & $[31.0,24.7]$ & 216.7 & $3.89\times10^{-11}$
& $1.20\times10^{-10}$& 1.05 \\
&Fig.~\ref{fig:3205DB}&
242.9 & 233.2 &($^{\rm c}$)  &($^{\rm c}$)  & 226.2 & $4.24\times10^{-11}$
& $1.20\times10^{-10}$ & 1.05\\
\hline
NGC\,1705 & Fig.~\ref{fig:1705}
& 73.44 & 73.31 & $3.33\times10^{19}$ & $[0.0962,0.287]$ & 73.31 & $2.89\times10^{-11}$
& $4.50\times10^{-10}$ & $\approx 1.00$\\
\hline
\end{tabular}
($^{\rm a}$): Coefficient $A_{\rm f} = 9.94 \times 10^{19}\kg\s^4\m^{-4}$
in equation~(\ref{eqn:BTFR}) \citep{McG08,McG12};
($^{\rm b}$): Exponent of declining RCs;\\ ($^{\rm c}$): Unrealistic test.
\label{tab:velocities}
\end{table*}

As a summary of the evaluations of the five galaxies, we compile in
Table~\ref{tab:masses} all mass values found in the literature and those
obtained through the various fits. Uncertainty ranges are not included,
but the large variations both of the published and the deduced values
indicate significant uncertainties. In Table~\ref{tab:velocities} the
observed and derived velocities of the RCs are listed together with
the fit parameter $g_{\rm fit}$ and the exponent $\beta$.

%
\section{Discussion and conclusions} 
\label{s:discuss}

With the exception of the irregular galaxy NGC\,1705, the other
acceleration diagrams in the lower panels indicate that for a good fit an
amplification factor~$F$ between 1.5 and 1.8 was required to raise the
$g_{\rm bar}$ values to $g_{\rm obs}$. According to
equation~({\ref{equ:lambda}}) the corresponding galactic radius has to be
$R_{\rm fit} \approx \lambda_0$ provided
most of the baryonic mass is within this limit.
Equations~(\ref{eqn:centripetal}), (\ref{equ:lambda}) and (\ref{eqn:fit})
then give for the flat RCs with $V_{\rm f}$ an estimate of
$A_{\rm fit} \approx 1/(g_{\rm fit}\,G_{\rm N})$
for the coefficients in equation~(\ref{eqn:BTFR}) and the corresponding
baryonic masses~$M_{\rm fit}$ in Table~\ref{tab:velocities}.
A comparison with mass values in Table~\ref{tab:masses} shows that
the masses are in reasonable agreement for NGC\,3198, NGC\,2404, UGC\,3205,
and even for NGC\,1705, but not for NGC\,1090.

The observed non-Keplerian behaviour of RCs could be modelled
for five disk galaxies without difficulties under the assumption of
multiple interactions of gravitons in the planes of the disks.
This can be considered as a substantial support for the
graviton impact theory\,--\,a modification of the old impact theory
proposed by Nicolas Fatio de Duillier in 1690.

As mentioned in Subsection~\ref{ss.NGC_3198}, the physics of the extraplanar
gas observations needs to be studied. This can only be done in a very
crude approximation at this stage. It will be helpful to visualize the
gravitational potential at $R > \lambda_0 \approx R_{\rm fit}$ both for the
case of Newton's law outside the galactic plane (a) and near that plane (b).
The situation will be considered explicitly at four radii
$R = R_{\rm fit} \times [1,2,3,4]$.
Case (a) gives the usual gravitational potential of
%
\begin{equation}
U(R) = -\frac{G_{\rm N}\,M_{\rm bar}}{R}
\end{equation}
\label{eqn:Newton}
with $U_0 = 0$ at $\infty$.
In case (b), we cannot assume a perfectly flat case, corresponding to
$g(R) \propto  1/R$, because this would lead
to $U_{\rm plane}(R) = \ln(R)$, which is not zero at infinity. However,
$g_{\rm mod}(R) \propto 1/R^\beta$ with $1 < \beta \ll 2$ is a good
approximation of the acceleration in the plane and can be integrated to
give the potential
%
\begin{equation}
U_{\rm mod}(R) = -
F\,\frac{G_{\rm N}\,M_{\rm bar}}
{(\beta-1)\,R_{\rm fit}}\left(\frac{R_{\rm fit}}{R}\right)^{(\beta - 1)}
\end{equation}
\label{eqn:plane}
when we require an amplification factor $F$ at $R_{\rm fit}$. The shape of
the potential near
the galactic plane can be compared to a canyon, if $\beta$ is increased
to 2 just above the plane.

We will apply these results to NGC\,3198 in Fig.~\ref{fig:3198} and
get with a baryonic mass of
$M_{\rm bar} = 6.91 \times 10^{40}\kg$ (i.e.~$3.46 \times 10^{10}\M$)
the accelerations indicated by the triangle marks and with
equation~(\ref{eqn:centripetal}) the velocities marked
in the upper panel, where the slow speeds correspond to the Keplerian case
outside the plane. The
ratio of the potentials is $U_{\rm mod}(R_{\rm fit})/U(R_{\rm fit}) = 35$,
i.e.
the ``depth'' of the canyon at $R_{\rm fit}$. The extraplanar material cannot
directly ``fall'' into the canyon, as this would violate the
angular momentum conservation (assuming a closed system). Energy and angular
momentum can, however, be conserved, if the falling material moves inwards
at the same time, because the radius decreases and the speed is nearly
constant.

Our model thus provides consistent pictures both for the in plane and
the extraplanar material.
Nevertheless, many questions remain, in particular, on the physical
properties of the gravitons. Since it was possible to explain the anomalous
RCs without assuming any DM that is also a complete mystery, there is hope
that further studies will clarify the situation. Considering that
MOND models are good fits for many galaxies, the multiple interaction
might even point to a physical process behind the mathematical modification
of the gravitational acceleration at small values.

\section*{Acknowledgements}
This research has made extensive use of the Astrophysics Data System (ADS).
Administrative support has been provided by the Max-Planck-Institute for
Solar System Research and the Indian Institute of Technology (Banaras Hindu
University).
We thank the anonymous referee for his/her constructive
and valuable comments
which improved the presentation of the manuscript.

\newpage


\bsp
\label{lastpage}

\begin{thebibliography}{00}
%
\bibitem[\protect\citeauthoryear{van Albada et al.}{1985}]{Albetal}
van Albada, T.\,S., Bahcall, J.\,N., Begeman, K., Sancisi, R., 1985,
Distribution of dark matter in the spiral galaxy NGC\,3198,
\apj, 295, 305
%
\bibitem[\protect\citeauthoryear{de Almeida, Amendola \& Niro}{2017}]{AlAmNi}
de Almeida, \'A., Amendola, L., Niro, V., 2017,
Galaxy rotation curves in modified gravity models,
J. Cosmology Astroparticle Phys., Issue~08, id. 012
%
%
\bibitem[\protect\citeauthoryear{Angus, van der Heyden \& Diaferio}{2012}]
{AnHeDi} 
Angus, G.\,W., van der Heyden, K.\,J., Diaferio, A., 2012,
The dynamics of the bulge dominated galaxy NGC\,7814 in MOND,
\aap, 543, id. A76
%
\bibitem[\protect\citeauthoryear{Annibali et al}{2003}]{Annetal}
Annibali, F., Greggio, L., Tosi, M., Aloisi, A., Leitherer, C., 2003,
The star formation history of NGC\,1705:
A poststarburst galaxy on the verge of activity,
\aj, 126, Issue~6, 2752
%
\bibitem[\protect\citeauthoryear{Babcock}{1939}]{Bab39}
Babcock, H.\,W., 1939,
The rotation of the Andromeda Nebula,
Lick Observatory Bull., 498, 41
%
%
\bibitem[\protect\citeauthoryear{Battaglieri et al.}{2017}]{Batetal}
Battaglieri, M., Belloni, A., Chou, A., and 248 authors, 2017,
US Cosmic Visions: New ideas in Dark Matter 2017: Community Report,
arXiv170704591
%
\bibitem[\protect\citeauthoryear{Begeman}{1987}]{Beg87}
Begeman, K.\,G., 1987,
HI rotation curves of spiral galaxies,
Ph.D. thesis, Kapteyn Institute
%
\bibitem[\protect\citeauthoryear{Begeman}{1989}]{Beg89}
Begeman, K.\,G., 1989,
HI rotation curves of spiral galaxies. I NGC\,3198,
\aap, 223, 47
%
\bibitem[\protect\citeauthoryear{Behroozi et al.}{2019}]{Behetal}
Behroozi, P., Becker, M.,  van den Bosch, F.\,C. and 20 authors, 2019,
Empirically constraining galaxy evolution,
BAAS, 51, Issue 3, id. 125
%
\bibitem[\protect\citeauthoryear{Bekenstein}{2010}]{Bek10}
Bekenstein, J.\,D., 2010,
Alternatives to dark matter:
Modified gravity as an alternative to dark matter,
arXiv:1001.3876
%
\bibitem[\protect\citeauthoryear{Binney, May \& Ostriker}{1987}]{BiMaOs}
Binney, J., May, A., Ostriker, J.\,P., 1987,
On the flattening of dark haloes,
\mnras, 226, 149
%
\bibitem[\protect\citeauthoryear{Blais-Ouellette, Amram \& Carignan}{2001}]
{BlAmCa}
Blais-Ouellette, S., Amram, P., Carignan, C., 2001,
Accurate determination of the mass distribution in spiral galaxies.
II. Testing the shape of dark halos ,
\aj, 121, Issue~4, 1952
%
\bibitem[\protect\citeauthoryear{de Blok}{2018}]{deB18}
de Blok, W.\,J.\,G., 2018,
Is there a universal alternative to dark matter?
Nature Astron., 2, 615
%
\bibitem[\protect\citeauthoryear{de Blok et al.}{2008}]{deB08}
de Blok, W.\,J.\,G.,  Walter, F., Brinks, E., Trachternach, C.,\\
Oh, S.-H., Kennicutt, R.\,C., Jr., 2008,
High-resolution rotation curves and galaxy mass models from THINGS ,
\aj, 136, Issue 6, 2648
%
\bibitem[\protect\citeauthoryear{de Blok et al.}{2014}]{deBetal}
de Blok, W.\,J.\,G., Keating, K.\,M., Pisano, D.\,J., Fraternali, F.,
Walter, F., Oosterloo, T., Brinks, E., Bigiel, F., Leroy, A., 2014,
A low H I column density filament in NGC\,2403:
signature of interaction or accretion ,
\aap, 569, id. A68
%
\bibitem[\protect\citeauthoryear{Bopp}{1929}]{Bop29}
Bopp, K., (Ed.), 1929,
Nicolas Fatio de Duillier: De la cause de la pesanteur,
Schriften der Stra{\ss}burger Wiss. Ges.\\ Heidelberg, 10, 19
%
\bibitem[\protect\citeauthoryear{Bottema \& Pesta\~{n}a}{2015}]{BotPes}
Bottema, R., Pesta\~{n}a, J.\,L.\,G., 2015,
The distribution of dark and luminous matter inferred from extended
rotation curves,
\mnras, 448, 2566
%
\bibitem[\protect\citeauthoryear{Boylan-Kolchin, Bullock \& Kaplinghat}
{2011}]{BoBuKa}
Boylan-Kolchin, M., Bullock, J.\,S., Kaplinghat, M., 2011,\\
Too big to fail?  The puzzling darkness
of massive milky way subhaloes,
\mnras, 415, L40
%
\bibitem[\protect\citeauthoryear{Bratton}{2011}]{Bra11}
Bratton, M., 2011,
The complete guide to the Herschel objects,
Cambridge University Press, Cambridge, UK
%
\bibitem[\protect\citeauthoryear{Burkert}{1995}]{Bur95}
Burkert, A., 1995,
The structure of dark matter halos in dwarf galaxies,
\apj, 447, L25
%
\bibitem[\protect\citeauthoryear{Daod \& Zeki}{2019}]{DaoZek}
Daod, N.\,A., Zeki, M.\,K., 2019,
Density and mass distribution of spiral galaxy NGC\,3198,
\apj, 870, id. 107,4
%
\bibitem[\protect\citeauthoryear{Desmond}{2017}]{Des17}
Desmond, H., 2017,
A statistical investigation of the mass discrepancy-acceleration relation,
\mnras, 464, Issue~4, 4160
%
\bibitem[\protect\citeauthoryear{Donato et al.}{2009}]{Donetal}
Donato, F., Gentile, G., Salucci, P., Frigerio Martins, C.,
Wilkinson, M.\,I., Gilmore, G., Grebel, E.\,K.; Koch, A., Wyse, R., 2009,
A constant dark matter halo surface density in galaxies,
\mnras, 397, Issue~3, 1169
%
\bibitem[\protect\citeauthoryear{Drew et al.}{2018}]{Dreetal}
Drew, P.\,M., Casey, C.\,M., Burnham, A.\,D., Hung, C.-L.,\\
Kassin, S\,A., Simons, R.\,C., Zavala, J.\,A., 2018,
Evidence of a flat outer rotation curve in a
star-bursting disk galaxy at\\ z = 1.6, \apj, 869, Issue~1, id. 58
%
\bibitem[\protect\citeauthoryear{Dutton et al.}{2019}]{Dutetal}
Dutton, A.\,A., Macci\`o, A.\.V., Obreja, A., Buck, T., 2019,
NIHAO - XVIII. Origin of the MOND phenomenology of galactic rotation curves
in a $\Lambda$CDM Universe,
\mnras, 485, Issue~2, 1886
%
\bibitem[\protect\citeauthoryear{Dyson et al.}{1920}]{Dys20}
Dyson, F.\,W., Eddington, A.\,S., Davidson, C., 1920.
A determination of the deflection of light by the Sun's gravitational field,
from observations made at the total eclipse of May 29, 1919,
Phil. Trans. R. astr. Soc. Lond. A, 220, 291
%
\bibitem[\protect\citeauthoryear{Eby et al.}{2016}]{Ebyetal}
Eby, J., Kouvaris, C., Nielsen, N.\,G., Gr{\o}nlund, N.,
Wijewardhana,\,L.\,C.\,R., 2016,
Boson stars from self-interacting dark matter,
J.\,High Ener.\,Phys., 2016, id.28
%
\bibitem[\protect\citeauthoryear{Einstein}{1916}]{Ein16}
Einstein, A., 1916,
Die Grundlage der allgemeinen Relativit\"atstheorie,
Ann. Phys. (Leipzig), 354, Issue 7, 769
%
\bibitem[\protect\citeauthoryear{Fatio de Duilleir}{1690}]{Fat90}
Fatio de Duilleir, N., 1690,
De la cause de la pesanteur.
Not. Rec. Roy. Soc. London, 6, 2, 125
%
%
\bibitem[\protect\citeauthoryear{Fraternali et al.}{2010}]{Fra10}
Fraternali, F., 2010,
Gas circulation and galaxy evolution ,
AIP Conf. Proc., 1240, 135
%
\bibitem[\protect\citeauthoryear{Fraternali \& Binney}{2006}]{FraBin06}
Fraternali, F., Binney, J.\,J., 2006,
A dynamical model for the extraplanar gas in spiral galaxies,
\mnras, 366, Issue 2, 449
%
\bibitem[\protect\citeauthoryear{Fraternali \& Binney}{2008}]{FraBin08}
Fraternali, F., Binney, J.\,J., 2008,
Accretion of gas on to nearby spiral galaxies,
\mnras, 386, Issue 2, 935
%
\bibitem[\protect\citeauthoryear{Fraternali et al.}{2001}]{Fraetal}
Fraternali, F., Oosterloo, T., Sancisi, R., van Moorsel, G., 2001,
A new, kinematically anomalous H I component in the spiral galaxy NGC\,2403,
\apj, 562, Issue 1, L47
%
\bibitem[\protect\citeauthoryear{Fraternali et al.}{2002}]{Fraetal02}
Fraternali, F., van Moorsel G., Sancisi, R., Oosterloo, T., 2002,
The HI halo of NGC\,2403,
\aj, 123, 3124
%
\bibitem[\protect\citeauthoryear{Fraternali, Oosterloo \& Sancisi}{2004}]{FraOosSan}
Fraternali, F., Oosterloo, T., Sancisi, R., 2004,
Kinematics of the ionised gas in the spiral galaxy NGC\,2403,
\aap, 424, 485
%
\bibitem[\protect\citeauthoryear{Fraternali, Sancisi \& Kamphuis}{2011}]
{FrSaKa}
Fraternali, F., Sancisi, R., Kamphuis, P., 2011,
A tale of two galaxies: Light and mass in NGC\,891 and NGC\,7814,
\aap, 531, A64
%
\bibitem[\protect\citeauthoryear{Freeman}{1970}]{Fre70}
Freeman, K.\,C., 1970,
On the disks of spiral and S0 galaxies,
\apj, 160, 811
%
\bibitem[\protect\citeauthoryear{Freeman}{2008}]{Fre08}
Freeman, K.\,C., 2008,
Formation and evolution of galaxy disks: Overview,
ASP Conf. Ser. 396, 3
%
\bibitem[\protect\citeauthoryear{Frigerio Martins \& Salucci}{2007}]{FriMSal}
Frigerio Martins, C., Salucci, P., 2007,
Analysis of rotation curves in the framework of $R^n$ gravity,
\mnras, 381, 1103
%
\bibitem[\protect\citeauthoryear{Gentile et al.}{2004}]{Geetal04}
Gentile, G., Salucci, P., Klein, U., Vergani, D., Kalberla, P., 2004,
The cored distribution of dark matter in spiral galaxies,
\mnras, 351, Issue 3, 903
%
\bibitem[\protect\citeauthoryear{Gentile et al.}{2009}]{Genetal}
Gentile, G., Famaey, B., Zhao, H., Salucci, P., 2009,
Universality of galactic surface densities within one dark halo
scale-length,
Nature, 461, 7264, 627
%
\bibitem[\protect\citeauthoryear{Gentile et al.}{2013}]{Genetal13}
Gentile, G., J\'ozsa, G.\,I.\,G., Serra, P., Heald, G.\,H.,\\
de Blok, W.\,J.\,G., Fraternali, F., Patterson, M.\,T.,\\
Walterbos, R.\,A.\, M., Oosterloo, T., 2013,
HALOGAS: Extraplanar gas in NGC\,3198,
\aap, 554, id.A125
%
\bibitem[\protect\citeauthoryear{Genzel et al.}{2017}]{Genetal17}
Genzel, R., F\"orster Schreiber, N.\,M., \"Ubler, H. and 18 authors, 2017,
Strongly baryon-dominated disk galaxies at the peak of galaxy
formation ten billion years ago,
Nature, 543, Issue 7645, 397
%
\bibitem[\protect\citeauthoryear{Ghari, Haghi \& Zonoozi}{2019}]{GhHaZo}
Ghari, A., Haghi, H., Zonoozi, A.\,H., 2019,
The radial acceleration relation and dark baryons in MOND,
\mnras, 487, 2, 2148
%
\bibitem[\protect\citeauthoryear{Ghari et al.}{2019}]{Ghaetal}
Ghari, A., Famaey, B., Laporte, C., Haghi, H., 2019,
Dark matter-baryon scaling relations from Einasto halo fits to
SPARC galaxy rotation curves ,
\aap, 623 id. A123
%
\bibitem[\protect\citeauthoryear{Giraud}{2000}]{Gir00}
Giraud, E., 2000,
A universal coupling relation between luminous and dark matter surface
densities in disk rotating galaxies,
\apj, 531, 701
%
\bibitem[\protect\citeauthoryear{Hague \& Wilkinson}{2014}]{HagWil}
Hague, P.\,R., Wilkinson, M.\,I., 2014,
Dark matter in disc galaxies -- II.
Density profiles as constraints on feedback scenarios,
\mnras, 443, 3712
%
\bibitem[\protect\citeauthoryear{Ja{\l}ocha et al.}{2011}]{Jaetal11}
Ja{\l}ocha, J., Bratek, {\L}., Kutschera, M., Skindzier, P., 2011,\\
Vertical gradients of azimuthal velocity in a global thin-disc model
of spiral galaxies NGC\,2403, NGC\,4559, NGC\,4302 and NGC\,5775,
\mnras, 412, 331
%
\bibitem[\protect\citeauthoryear{Jarrett et al.}{2003}]{Jaretal}
Jarrett, T.\,H., Chester, T., Cutri, R., Schneider, S.\,E.,\\
Huchra, J.\,P., 2003,
The 2MASS large galaxy atlas,
\aj, 125, 525
%
\bibitem[\protect\citeauthoryear{Kaplinghat, Ren \& Yu}{2019}]{KaReYu}
Kaplinghat, M., Ren, T., Yu, H.-B., 2019,
Dark matter cores and cusps in spiral galaxies and their explanations,
arXiv:1911.00544
%
\bibitem[\protect\citeauthoryear{Karukes, Salucci \& Gentile}{2015}]{KaSaGe}
Karukes, E.\,V., Salucci, P., Gentile, G., 2015,
The dark matter distribution in the spiral NGC\,3198 out to 0.22 $R_{\rm vir}$,
\aap, 578, id.A13
%
\bibitem[\protect\citeauthoryear{Karukes \& Salucci}{2017}]{KarSal}
Karukes, E.\,V., Salucci, P., 2017,
The universal rotation curve of dwarf disc galaxies,
\mnras, 465, 4703
%
\bibitem[\protect\citeauthoryear{Kassin, de Jong \& Weiner}{2006}]{KadJoWei}
Kassin, S.\,A., de Jong, R.\,S., Weiner, B.\,J., 2006,
Dark and baryonic matter in bright spiral galaxies. II.
Radial distributions for 34 galaxies ,
\apj, 643, 804 (attn.: table 2 of arXiv:astro-ph/0602027v1)
%
\bibitem[\protect\citeauthoryear{Kostov}{2006}]{Kos06}
Kostov, V., 2006,
Mass distribution of spiral galaxies in a thin disk model with velocity
curve extrapolation,
arXiv:astro-ph/0604395
%
\bibitem[\protect\citeauthoryear{Kregel, van der Kruit \& de Grijs}{2002}]{KrKrGr}
Kregel, M., van der Kruit, P.\,C., de Grijs, R., 2002,\\
Flattening and truncation of stellar discs in edge-on spiral galaxies,
\mnras, 334, 646
%
\bibitem[\protect\citeauthoryear{Kroupa, Pawlowski \& Milgrom}{2012}]{KrPaMi}
Kroupa, P., Pawlowski, M., Milgrom, M., 2012,\\
The failures of the Standard Model of Cosmology require a new paradigm,
Int. J. Mod. Phys. D, 21, id. 1230003
%
\bibitem[\protect\citeauthoryear{von Laue}{1959}]{Lau59}
von Laue, M., 1959,
Geschichte der Physik, 4. erw. Aufl.
Ullstein Taschenb\"ucher-Verlag, Frankfurt/Main
%
\bibitem[\protect\citeauthoryear{Lelli et al.}{2017}]{Leletal}
Lelli, F., McGaugh, S.\,S., Schombert, J.\,M., Pawlowski, M.\,S., 2017,
One law to rule them all: The radial acceleration relation of galaxies,
\apj, 836, Issue 2, id. 152
%
\bibitem[\protect\citeauthoryear{Li, Tang \& Lin}{2017}]{LiTaLi}
Li, X., Tang, L., Lin, H.-N., 2017,
Comparing dark matter models, modified Newtonian dynamics and
modified gravity in accounting for galaxy rotation curves,
Chin. Phys. C, 41, Issue~5, id. 055101
%
\bibitem[\protect\citeauthoryear{Li et al.}{2019a}]{Lietala}
Li, P., Lelli, F., McGaugh, S., Pawlowski, M.\,S.,
Zwaan, M.\,A., Schombert, J., 2019a,
The halo mass function of late-type galaxies from H I kinematics,
\apj, 886,1, id L11
%
\bibitem[\protect\citeauthoryear{Li et al.}{2019b}]{Lietalb}
Li, P., Lelli, F., McGaugh, S., Starkman, N., Schombert, J., 2019b,
A constant characteristic volume density of dark matter haloes
from SPARC rotation curve fits,
\mnras, 482, Issue 4, 5106
%
\bibitem[\protect\citeauthoryear{Lovas \& Kielkopf}{2014}]{LovKie}
Lovas, S., Kielkopf, J.\,F., 2014,
Distribution of dark and luminous mass in galaxies,
\aj, 147, id. 135,5
%
\bibitem[\protect\citeauthoryear{McGaugh}{2005a}]{McG05a}
McGaugh, S.\,S., 2005a,
Balance of dark and luminous mass in rotating galaxies,
\prl, 95, id. 171302
%
\bibitem[\protect\citeauthoryear{McGaugh}{2005b}]{McG05b}
McGaugh, S.\,S., 2005b,
The baryonic Tully-Fisher relation of galaxies with extended rotation curves
and the stellar mass of rotating galaxies,
\apj, 632, 859
%
\bibitem[\protect\citeauthoryear{McGaugh}{2008}]{McG08}
McGaugh, S.\,S., 2008,
Balance of dark and luminous mass in rotating galaxies,
\prl, 95, Issuee 17, id. 171302
%
\bibitem[\protect\citeauthoryear{McGaugh}{2012}]{McG12}
McGaugh, S.\,S., 2012,
The baryonic Tully-Fisher relation of gas-rich galaxies as a test of
$\Lambda$CDM and MOND,\\
\aj, 143, id. 40
%
\bibitem[\protect\citeauthoryear{McGaugh}{2014}]{McG14}
McGaugh, S.\,S., 2014,
The third law of galactic rotation ,
Galaxies, 2, 601
%
\bibitem[\protect\citeauthoryear{McGaugh, Lelli \& Schombert}{2016}]{McLeSc}
McGaugh, S.\,S., Lelli, F., Schombert, J.\,M., 2016,\\
Radial acceleration relation in rotationally supported galaxies,
\prl, 117, id. 201101
%
\bibitem[\protect\citeauthoryear{McGaugh et al.}{2019}]{McGetal}
McGaugh, S.\,S., Lelli, F., Li, P., Schombert, J.\,M., 2019,
Dynamical regularities in galaxies,
arXiv:1909.02011
%
\bibitem[\protect\citeauthoryear{Milgrom}{1983}]{Mil83}
Milgrom, M., 1983,
A modification of the Newtonian dynamics as a possible alternative to the
hidden mass hypothesis
\apj, 270, 365
%
\bibitem[\protect\citeauthoryear{Milgrom}{1994}]{Mil94}
Milgrom, M., 1994,
Dynamics with a nonstandard inertia-ac\-cel\-er\-a\-tion relation:
An alternative to dark matter in galactic systems,
Ann. Phys., 229, 384
%
\bibitem[\protect\citeauthoryear{Milgrom}{2015}]{Mil15}
Milgrom, M., 2015,
MOND theory,
Can. J. Phys., 93, 107
%
\bibitem[\protect\citeauthoryear{Milgrom}{2016}]{Mil16}
Milgrom, M., 2016,
MOND impact on and of the recently
updated mass-discrepancy-acceleration relation,
arXiv:1609.06642
%
\bibitem[\protect\citeauthoryear{Milgrom}{2020}]{Mil20}
Milgrom, M., 2020,
The $a_0$\,--\,cosmology connection in MOND,
arXiv:2001.09729
%
\bibitem[\protect\citeauthoryear{Milgrom \& Sanders}{2007}]{MilSan}
Milgrom, M., Sanders, R.\,H., 2007,
Modified Newtonian Dynamics Rotation curves of very low mass spiral galaxies,
\apj, 658, Issue~1, L17
%
\bibitem[\protect\citeauthoryear{Navarro}{1998}]{Nav98}
Navarro, J.\,F., 1998,
The cosmological significance of disk galaxy rotation curves,
arXiv:astro-ph/9807084
%
\bibitem[\protect\citeauthoryear{Navarro et al.}{2017}]{Navetal}
Navarro, J.\,F., Ben\'itez-Llambay, A., Fattahi, A.,
Frenk, C.\,S., Ludlow, A.\,D., Oman, K.\,A., Schaller, M., Theuns, T., 2017,
The origin of the mass discrepancy-acceleration relation in $\Lambda$CDM ,
\mnras, 471, Issue 2, 1841
%
\bibitem[\protect\citeauthoryear{Nijhoff}{1901}]{Nij01}
Nijhoff, M., (Ed.), 1901,
Correspondance, 1685--1690, Nos. 2570 et 2582.
{\OE}uvres Compl\`{e}tes de Christiaan Huygens, IX, 381 and 407, La Haya
%
\bibitem[\protect\citeauthoryear{Noordermeer et al.}{2007}]{Nooetal}
Noordermeer, E., van der Hulst, J.\,M., Sancisi, R., Swaters, R.\,S.,
van Albada, T.\,S., 2007,
The mass distribution in early-type disc galaxies: declining rotation curves
and correlations with optical properties,
\mnras, 376, Issue~4, 1513
%
\bibitem[\protect\citeauthoryear{O'Halloran et al.}{2010}]{OHaetal}
O'Halloran, B., Galametz, M., Madden, S.\,C. and 56 authors, 2010,
Herschel photometric observations of the low metallicity
dwarf galaxy NGC\,1705,
\aap, 518, id. L58
%
\bibitem[\protect\citeauthoryear{O'Meara}{2011}]{OMea}
O'Meara, S.\,J., 2011,
Deep-sky companions: The secret deep,
Cambridge University Press, Cambridge, UK
%
\bibitem[\protect\citeauthoryear{Oort}{1932}]{Oor32}
Oort, J.\,H., 1932,
The force exerted by the stellar system in the direction perpendicular
to the galactic plane and some related problems,
Bull. Astron. Inst. Netherlands, 6, 249
%
\bibitem[\protect\citeauthoryear{Oort}{1940}]{Oor40}
Oort, J.\,H., 1940,
Some problems concerning the structure and dynamics of the
galactic system and the elliptical nebulae NGC\,3115 and 4494,
\apj, 91, 273
%
\bibitem[\protect\citeauthoryear{ Persic \& Salucci}{1990}]{PerSal}
Persic, M., Salucci, P., 1990,
The disc contribution to rotation curves of spiral galaxies,
\mnras, 247, 349
%
\bibitem[\protect\citeauthoryear{ Persic, Salucci \& Stel}{1996}]{PeSaSt}
Persic, M., Salucci, P., Stel, F., 1996,
The universal rotation curve of spiral galaxies -- I. The dark matter connection,
\mnras, 281, Issue~1, 27 (Note Erratum: \mnras, 283, 1102)
%
\bibitem[\protect\citeauthoryear{Posti et al.}{2019}]{Posetal}
Posti, L., Marasco, A., Fraternali, F., Famaey, B., 2019,
Galaxy disc scaling relations:
A tight linear galaxy-halo connection challenges abundance matching,
\aap, 629, id. A59
%
\bibitem[\protect\citeauthoryear{Richards et al.}{2018}]{Ricetal}
Richards, E.\,E., van Zee, L., Barnes, K.\,L., Staudaher, S.,\\ Dale, D.\,A.,
Braun, T.\,T., Wavle, D.\,C., Dalcanton, J.\,J., Bullock, J.\,S.,
Chandar, R., 2018,
Baryonic distributions in galaxy dark matter haloes - II. Final results .
\mnras, 476, Issue 4, 5127
%
\bibitem[\protect\citeauthoryear{Rodrigues et al.}{2018}]{Rodetal}
Rodrigues, D.\,C., Marra, V., del Popolo, A., Davari, Z., 2018,
Absence of a fundamental acceleration scale in galaxies,
Nature Astron., 2, 668
%
%
\bibitem[\protect\citeauthoryear{Rubin}{1983}]{Rub83}
Rubin, V.\,C., 1983,
Dark matter in spiral galaxies,
Scientific American, 248, 96
%
\bibitem[\protect\citeauthoryear{Rubin}{1986}]{Rub86}
Rubin, V.\,C., 1986,
Dark matter in the Universe,
Highlights of Astronomy, 7, 27
%
\bibitem[\protect\citeauthoryear{Rubin}{2000}]{Rub00}
Rubin, V.\,C., 2000,
One hundred years of rotating galaxies,
\pasp, 112, 747
%
\bibitem[\protect\citeauthoryear{Salucci}{2018}]{Sal18}
Salucci, P., 2018,
Dark Matter in galaxies: Evidences and challenges,
Found. Phys. 48, Issue~10, 1517
%
\bibitem[\protect\citeauthoryear{Salucci}{2019}]{Sal19}
Salucci, P., 2019,
The distribution of dark matter in galaxies,
Astron. Astrophys. Rev., 27, Issue~1, id. 2
%
\bibitem[\protect\citeauthoryear{Salucci \& Turini}{2017}]{SalTur}
Salucci, P., Turini, N., 2017,
Evidences for collisional dark matter in galaxies?
arXiv:1707.01059
%
\bibitem[\protect\citeauthoryear{Samurovi\'{c}}{2016}]{Sam16}
Samurovi\'{c}, S., 2016,
The Newtonian and MOND dynamical models of NGC\,5128: Investigation of the
dark matter contribution,
Serbian Astron. J., 192, 9
%
\bibitem[\protect\citeauthoryear{Sancisi}{2004}]{Sanc04}
Sancisi, R., 2004,
The visible matter -- dark matter coupling,
IAUS, 220, 233
%
\bibitem[\protect\citeauthoryear{Sanders}{2019}]{San19}
Sanders, R.\,H., 2019,
The prediction of rotation curves in gas-dominated dwarf galaxies
with modified dynamics,
\mnras, 485, Issue 1, 513
%
\bibitem[\protect\citeauthoryear{Sanders \& Noordermeer}{2007}]{SanNoo}
Sanders, R.\,H., Noordermeer, E., 2007,
Confrontation of MOdified Newtonian Dynamics with the rotation curves
of early-type disc galaxies,
\mnras, 379, 702
%
\bibitem[\protect\citeauthoryear{Santos-Santos et al.}{2019}]{Sanetal}
Santos-Santos, I.\,M.\,E., Navarro, J.\,F., Robertson, A.,
Ben\'itez-Llambay, A., Oman, K.\,A., Lovell, M.\,R.,
Frenk, C.\,S.,\\ Ludlow, A.\,D., Fattahi, A., Ritz, A., 2019,
Baryonic clues to the puzzling diversity of dwarf galaxy rotation curves,
arXiv:1911.09116
%
\bibitem[\protect\citeauthoryear{Schaap, Sancisi \& Swaters}{2000}]{SchSS}
Schaap, W.\,E., Sancisi, R., Swaters, R.\,A., 2000,
The vertical extent and kinematics of the HI in NGC\,2403,
\aap, 356, L49
%
\bibitem[\protect\citeauthoryear{Schneider et al.}{2017}]{Sch17}
Schneider, A., Trujillo-Gomez, S., Papastergis, E., Reed, D.\,S.,
Lake, G., 2017,
Hints against the cold and collisionless nature of dark matter from the
galaxy velocity function,
\mnras, 470, 1542
%
\bibitem[\protect\citeauthoryear{Sofue}{2016}]{Sof16}
Sofue, Y., 2016,
Rotation curve decomposition for size-mass relations of bulge, disk,
and dark halo components in spiral galaxies,
Publ. Astron. Soc. Japan, 68, Issue~1, id.2
%
\bibitem[\protect\citeauthoryear{Sofue \& Rubin}{2001}]{SofRub}
Sofue, Y., Rubin, V., 2001,
Rotation curves of spiral galaxies,
Ann. Rev. Astron. Astrophys., 39, 137
%
\bibitem[\protect\citeauthoryear{Soldner}{1804}]{Sol04}
Soldner, J., 1804,
Ueber die Ablenkung eines Lichtstrals von seiner geradlinigen Bewegung
durch die Attraktion eines Weltk\"orpers, an welchem er nahe vorbei geht,
Berliner Astron. Jahrb. 161
%
\bibitem[\protect\citeauthoryear{Stark, McGaugh \& Swaters}{2009}]{StMcSw}
Stark, D.\,V., McGaugh, S.\,S., Swaters, R.\,A., 2009,
A first attempt to calibrate the baryonic Tully-Fisher relation
with gas-dominated galaxies,
\aj, 138, Issue 2, 392
%
\bibitem[\protect\citeauthoryear{Taga \& Iye}{1994}]{TaIy}
Taga, M., Iye, M., 1994,
Halo model of spiral galaxy NGC\,3198,
\mnras, 271, 427
%
\bibitem[\protect\citeauthoryear{Tenneti et al.}{2018}]{Tenetal}
Tenneti, A., Mao, Y.-Y., Croft, R.\,A.\,C., Di Matteo, T.,\\
Kosowsky, A., Zago, F., Zentner, A.\,R., 2018,
The radial acceleration relation in disc galaxies in the
MassiveBlack-II simulation,
\mnras, 474, Issue 3, 3125
%
\bibitem[\protect\citeauthoryear{Tosi et al.}{2001}]{Tosetal}
Tosi, M., Sabbi, E., Bellazzini, M., Aloisi, A., Greggio, L.,\\
Leitherer, C., Montegriffo, P., 2001,
The resolved stellar populations in NGC\,1705,
\aj, 122, 1271
%
\bibitem[\protect\citeauthoryear{Treu \& Koopmans}{2004}]{TreKoo}
Treu, T., Koopmans, L.\,V.\,E., 2004,
Massive dark matter halos and evolution of early-type galaxies to
$z \approx 1$,
\apj, 611, 739
%
\bibitem[\protect\citeauthoryear{Tully \& Fisher}{1977}]{TulFis}
Tully, R.\,B., Fisher, J.\,R., 1977,
A new method of determining distances to galaxies,
\aap, 54 (3), 661
%
\bibitem[\protect\citeauthoryear{Tully, Courtois \& Sorce}{2016}]{TuCoSo}
Tully, R.\,B., Courtois, H.\,M., Sorce, J.\,G., 2016,
Cosmicflows-3 ,
\aj, 152, Issue 2, id. 50
%
%
\bibitem[\protect\citeauthoryear{Le Verrier}{1859}]{Ver59}
Le Verrier, U., 1859,
Lettre de M. Le Verrier \'a M.
Faye sur la th\'eorie de Mercure et sur le mouvement du p\,erih\'elie de
cette plan\`ete,
Compt. rend. hebdomad. s\'eanc. Acad. sci., 49, 379
%
%
\bibitem[\protect\citeauthoryear{Wechsler \& Tinker}{2018}]{WecTin}
Wechsler, R.\,H., Tinker, J.\,L., 2018,
The connection between galaxies and their Dark Matter halos,
Ann. Rev. Astron. Astrophys., 56, 435
%
\bibitem[\protect\citeauthoryear{Weinberg et al.}{2015}]{Weietal}
Weinberg,\,D.\,H., Bullock,\,J.\,S., Governato,\,F., Kuzio de Naray,\,R.,
Peter, A.\,H.\,G., 2015,
Cold dark matter: Controversies on small scales,
Proc. Nat. Acad. Sci., 112, 12249
%
\bibitem[\protect\citeauthoryear{Wilhelm, Wilhelm \& Dwivedi}{2013}]{WilWilDwi}
Wilhelm, K., Wilhelm, H., Dwivedi, B.\,N., 2013,
An impact model of Newton's law of gravitation,
\apss, 343, 135
%
\bibitem[\protect\citeauthoryear{Wilhelm \& Dwivedi}{2014}]{WilDwi}
Wilhelm, K., Dwivedi, B.\,N., 2014,
Secular perihelion advances of the inner planets and Asteroid Icarus,
\na, 31, 5
%
\bibitem[\protect\citeauthoryear{Wilhelm \& Dwivedi}{2015a}]{WilDwi15a}
Wilhelm, K., Dwivedi, B.\,N., 2015a,
On the potential energy in a gravitationally bound two-body system,
New Astron., 34, 250
%
\bibitem[\protect\citeauthoryear{Wilhelm \& Dwivedi}{2015b}]{WilDwi15b}
Wilhelm, K., Dwivedi, B.\,N., 2015b,
Anomalous Earth flybys of spacecraft,
\apss, 358, id.~18
%
\bibitem[\protect\citeauthoryear{Wilhelm \& Dwivedi}{2018}]{WiDwGal}
Wilhelm, K., Dwivedi, B.\,N., 2018,
A physical process of the radial acceleration of disc galaxies,
\mnras, 474, Issue 4, 4723
%
\bibitem[\protect\citeauthoryear{Wilhelm \& Dwivedi}{2020}]{WiDi20}
Wilhelm, K., Dwivedi, B.\,N., 2020,
Impact models of gravitational and electrostatic forces,
Chapter 5 in Planetology (ed.~B. Palaszewski), DOI: 10.5772/intechopen.75213
%
\bibitem[\protect\citeauthoryear{Zehe}{1983}]{Zeh83}
Zehe, H., 1983,
Die Gravitationstheorie des Nicolas Fatio de Duillier.
Arch. Hist. Exact Sci., 28, 1

\bibitem[\protect\citeauthoryear{Zwicky}{1933}]{Zwi33}
Zwicky, F., 1933,
Die Rotverschiebung von extragalaktischen Nebeln,
Helvetica Physica Acta, 6, 110
%

%
\end{thebibliography}
\end{document}